\documentclass[aps, prl, reprint, twocolumn, superscriptaddress, longbibliography]{revtex4-2}
\usepackage[utf8]{inputenc}
\usepackage[T1]{fontenc}
\usepackage{lmodern}
\usepackage{graphicx}
\usepackage{amsmath,amssymb}
\usepackage{xcolor}
\usepackage{booktabs}
\usepackage{tabularx}
\usepackage{caption}
\usepackage{subcaption}
\usepackage[colorlinks,allcolors=black,citecolor=blue,urlcolor=blue]{hyperref}
\usepackage[mathlines]{lineno}
\usepackage[normalem]{ulem}
\usepackage{placeins}
\usepackage{textcomp}
\usepackage{comment}
\graphicspath{{figures/}}

\DeclareUnicodeCharacter{2212}{-}
\begin{document}
\def\mytitle{From molecular to atomic superionic water under extreme confinement and pressure}
\title{\mytitle}

\author{Ioachim G. Dusa}
\email{igd27@cam.ac.uk}
\affiliation{Cavendish Laboratory, Department of Physics, University of Cambridge, Cambridge CB3 0HE, UK}
\affiliation{Lennard-Jones Centre, University of Cambridge, Trinity Lane, Cambridge CB2 1TN, UK}

\author{Chris J. Pickard}
\affiliation{Department of Materials Science \& Metallurgy, University of Cambridge, 27 Charles Babbage Road, Cambridge CB3 0FS, UK}
\affiliation{Lennard-Jones Centre, University of Cambridge, Trinity Lane, Cambridge CB2 1TN, UK}

\author{Angelos Michaelides}
\affiliation{Yusuf Hamied Department of Chemistry, University of Cambridge, Lensfield Road, Cambridge CB2 1EW, UK}
\affiliation{Lennard-Jones Centre, University of Cambridge, Trinity Lane, Cambridge CB2 1TN, UK}

\author{Christoph Schran}
\email{cs2121@cam.ac.uk}
\affiliation{Cavendish Laboratory, Department of Physics, University of Cambridge, Cambridge CB3 0HE, UK}
\affiliation{Lennard-Jones Centre, University of Cambridge, Trinity Lane, Cambridge CB2 1TN, UK}

\date{\today}

\begin{abstract}
Confined at the nanoscale, water forms phases with no bulk counterpart.
It turns superionic at
around 10~GPa, an order of magnitude below bulk, while preserving its molecular identity.
In contrast, bulk water becomes superionic and non-molecular only under extreme conditions, beyond 100~GPa and thousands of kelvin.
How these two superionic states connect, where confinement meets extreme pressure, remains unexplored.
Here, 
using
machine learning potentials at first-principles accuracy,
we map
the phase behaviour of
water confined between idealised walls from ambient conditions to 550~GPa.
We find that confined water hosts two distinct superionic states: a molecular one that persists up to 225 GPa, and an atomic one with symmetric \mbox{O--H--O} bonds at even higher pressures.
Across this crossover, proton transport changes from the Grotthuss mechanism to angular hops of the hydrogens about the oxygens.
Replacing the idealised walls with fully flexible diamond surfaces, we find that both transport regimes persist.
Electronically, the oxygen lone pairs survive compression and the band gap of the water--wall system falls well below that of bulk water.
%
Confinement thus opens a new route to high-pressure phases of water, reshaping its structure, proton dynamics and electronic properties alike.
\end{abstract}
\maketitle

\section{Introduction}
\label{sec:intro}

When water is confined to the nanoscale, its structure and dynamics can differ profoundly from its bulk behaviour.
In the low- to intermediate-pressure regime (1~bar -- 50~GPa), this has been established both experimentally and computationally:
Experiments on water confined between graphene and boron nitride sheets have revealed ferroelectric ordering in two-dimensional ice at ambient conditions~\cite{ferro}, nearly frictionless molecular flow through sub-nanometre channels~\cite{frictionless}, and a strongly anisotropic dielectric response, with components greatly suppressed or enhanced relative to the bulk value~\cite{Fumagalli,Fumagali2}.
Moreover, \textit{ab initio} simulations have shown that, already at low gigapascal pressures, confinement stabilises quasi-two-dimensional hydrogen-bond networks that differ qualitatively from bulk ice phases~\cite{2DIce-ab-init, ravindra_quasi-one-dimensional_2024}.
More recent studies have extended this picture using machine learning potentials (MLPs)~\cite{MLPBehlerReview, MLPGaborReview, MLPGroupReview} into the multi-GPa regime, revealing superionic behaviour, ultrafast one-dimensional proton hopping, and a sequence of structural transitions from monolayer to bilayer ice that have no bulk analogue~\cite{Kapil2022, jiang_rich_2024, coles2025nanoconfinedsuperionicwatermolecular, lin_temperature-pressure_2023}.
Notably, the confined superionic phase appears at pressures one order of magnitude below the bulk superionic transition~\cite{Kapil2022,jiang_rich_2024}, and, in sharp contrast to bulk superionic ice, it remains composed of intact water molecules, making it a molecular superionic phase~\cite{coles2025nanoconfinedsuperionicwatermolecular}.
Confinement can thus bring exotic high-pressure physics within accessible thermodynamic reach, while profoundly changing the properties of water.

Compression is itself one of the most powerful levers on the structure and dynamics of water.
In bulk ice, pressure progressively symmetrises the hydrogen bonds and absorbs the oxygen lone pairs into a symmetric bonding network~\cite{icexwannier}, until discrete molecules give way to the non-molecular ice X at roughly 120~GPa~\cite{highpressureice}.
The band gap nevertheless stays open up to terapascal pressures~\cite{highpressureice}, where water is predicted to decompose into hydrogen-rich phases~\cite{pickard_decomposition_2013}.
At elevated temperatures, extreme compression drives water into a superionic state, in which protons diffuse within a crystalline oxygen sublattice~\cite{millot_nanosecond_2019, noauthor_superionic_nodate} with protonic conductivity exceeding $0.1$ or $1~\mathrm{S\,cm^{-1}}$~\cite{Kapil2022, iceX-hopping}.
Crucially, these predictions of novel physics have been validated experimentally:
Shock compression has revealed superionic ice through nanosecond X-ray diffraction~\cite{millot_nanosecond_2019}, laser-heated diamond anvil cells have characterised superionic ice phases under static compression~\cite{natureDAC, prlDAC}, and diamond anvils now reach pressures of up to 500~GPa~\cite{jenei_single_2018}.
Comparable conditions prevail in the mantles of ice giants~\cite{InteriorOfGiantPlanets, REDMER2011798}, where water is also expected to occur at mineral interfaces rather than solely as a bulk fluid~\cite{chen_closest_packing_2020, kovacevic_miscibility_2022, pan_magnesium_2023}.

Confinement and extreme pressure have so far been studied largely in isolation.
While confined water has been mapped up to tens of gigapascals, the regime above 100~GPa, where the two meet, remains a no man's land.
This leaves open whether the molecular superionic state survives compression, whether the molecular-to-atomic transition occurs under confinement, and how proton transport changes across it.
These questions are
also in reach of experiment: diamond anvil cells compress samples between two flat diamond surfaces, and when paired with nanoscale hosts such as layered materials~\cite{chen_closest_packing_2020}, they can probe confined water at extreme pressures.
Such geometries also raise the question of the interaction with the walls themselves.
Implicit walls, smooth potentials replacing the confining material, offer an efficient and widely used description of confinement, while explicit atomistic walls have so far been limited to rigid setups or to graphene at a few gigapascals~\cite{Kapil2022, metallicHydrogen}.
It therefore remains open whether predictions from implicit walls hold for a real confining material at extreme pressure.

Addressing these questions reliably poses two main challenges.
On the one hand, the underlying electronic-structure calculations become increasingly demanding at high pressures, as compressed geometries make numerical convergence more delicate, rendering extended \textit{ab initio} molecular dynamics (MD) and structure searches computationally prohibitive.
On the other hand, conventional interatomic potentials struggle to remain accurate across the broad pressure range considered here, from ambient conditions to several hundred gigapascals.
MLPs provide a route around both limitations.
By replacing self-consistent-field electronic-structure evaluations, they enable large-scale simulations at accessible computational cost, while
retaining high
accuracy across strongly varying bonding environments.

Here, we go beyond previous work on confined water in three respects.
First, we extend the explored pressure range far beyond the tens of gigapascals studied so far
into this uncharted regime.
For that purpose, we train a MACE model~\cite{batatia2023macehigherorderequivariant,batzner_e3-equivariant_2022} on hybrid DFT reference data (revPBE0-D3), agreeing closely with quantum Monte Carlo reference calculations for confined water~\cite{Kapil2022}. 
With it, we map the phase behaviour within one consistent framework from ambient conditions to $\sim$550~GPa using random structure searches~\cite{AIRSS} and finite temperature molecular dynamics.
Second, since hydrogen--wall interactions can no longer be neglected at these pressures, we describe the confinement through an improved Lennard--Jones potential parametrised for water--graphene interactions~\cite{vekeman_development_2023}, with species-specific terms for oxygen and hydrogen.
Third, we go beyond implicit walls with an explicit model resembling a diamond anvil cell, in which water is confined between fully flexible (111) diamond surfaces terminated with fluorine~\cite{FterminatedD}.
We treat this realistic confinement model 
at the PBE level, both \textit{ab initio} and at scale with ephemeral data-derived potentials (EDDPs)~\cite{EDDPrandomstructure, EDDPlammps}, reaching thousands of atoms and nanosecond timescales.
Since the two setups differ in confinement model, interatomic potential and reference functional, their agreement provides an independent test of the physics reported here.
%

To reveal the signatures of confinement, we consider an extreme-confinement regime represented by a narrow \(5~\text{\AA}\) slit~\cite{Kapil2022}.
Probing the full pressure range within one consistent framework, we recover established low- to intermediate-pressure phases~\cite{Kapil2022, 2DIce-ab-init, jiang_rich_2024, lin_temperature-pressure_2023} and link them to the high-pressure regime.
Above 100\,GPa, we identify three previously unreported bilayer phases and a molecular-to-atomic transition, in which discrete water molecules give way to symmetric O--H--O bonds.
At elevated temperatures, confined water becomes superionic in both a molecular and an atomic form.
The molecular regime has no bulk counterpart and persists up to $\sim$225~GPa, extending molecular superionicity~\cite{coles2025nanoconfinedsuperionicwatermolecular} far beyond the pressures at which it was first predicted.
It transports protons by hopping along hydrogen bonds combined with rotations about the host oxygen, whereas the atomic regime transports them predominantly through angular hops around the oxygen sites.
The two are separated by a conductivity minimum at the molecular-to-atomic crossover.
Between explicit diamond walls, both regimes and a conductivity minimum between them persist, even though the oxygen lattice no longer undergoes any transition, pointing to a local origin of the crossover rather than one set by the crystal structure.

Electronically, confinement preserves the valence oxygen lone pairs, which in bulk ice are absorbed into the symmetric bonding network~\cite{icexwannier}.
We find that compression instead raises their band edge in energy under confinement.
Whether the band gap closes, however, is controlled by the confining material.
Between implicit walls, the gap closes at $\sim$500~GPa at the GGA level.
Between explicit diamond walls, the gap instead remains open at about 1.8~eV at 500~GPa, still 6.6~eV below that of compressed bulk ice~\cite{highpressureice} and 3.7~eV below that of diamond at the same pressure.
This follows from the band alignment between the water and the fluorine-terminated diamond, which places the carbon conduction states above the oxygen valence states but below the hydrogen-derived ones.

More broadly, our results demonstrate that extreme confinement reshapes bonding topology, proton transport and valence electronic structure under megabar compression, establishing nanoconfinement as a
new control parameter
for high-pressure physics.

\section{Results}
\label{sec:Results}
\subsection*{Structural evolution: from molecular to atomic nanoconfined ice}

\begin{figure*}[t]  
  \centering
  \includegraphics[width=\linewidth]{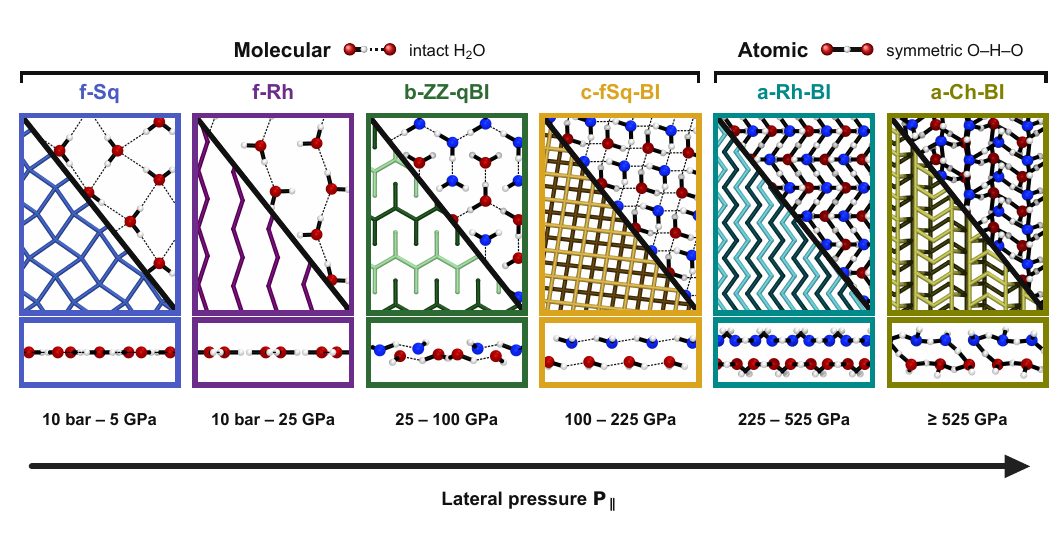}
\caption{\textbf{Structural evolution of nanoconfined ice under lateral pressure and implicit confinement.}
\\
Lowest-enthalpy $T = 0~\text{K}$ structures of water implicitly confined to a 5~\AA{} slit as a function of lateral pressure $P_\parallel$ (10~bar--550~GPa).
Each phase is shown in top view (upper box) and side view (lower box).
In the lower-left triangle of the top views, vertices denote oxygen atoms and edges denote O--H$\cdots$O hydrogen bonds between intact water molecules in the molecular phases, or symmetric O--H--O bonds in the atomic phases.
Light and dark shades distinguish sheets along the confinement direction ($\hat{z}$).
In the upper-right triangle, the phase is fully displayed with oxygens (red and blue, corresponding to different sheets) and hydrogens (white).
Dotted lines represent O--H$\cdots$O hydrogen bonds.
In the atomic phases (prefix a-), protons sit midway between neighbouring oxygens and discrete molecules no longer exist.
Where pressure intervals overlap, the phases are enthalpically degenerate to within 5~meV/water.
The sequence spans monolayer ice (flat-square and flat-rhombic, f-Sq/f-Rh)~\cite{2DIce-ab-init, Kapil2022}, quasi-bilayer ice (branched zig-zag, b-ZZ-qBI)~\cite{jiang_rich_2024, lin_temperature-pressure_2023}, and bilayer phases (centred flat-square c-fSq-BI, atomic rhombic a-Rh-BI, and atomic chevron a-Ch-BI).
}
\label{fig:structural-evolution}
\end{figure*}

\begin{figure*}[t]
  \centering
  \includegraphics[width=\textwidth]{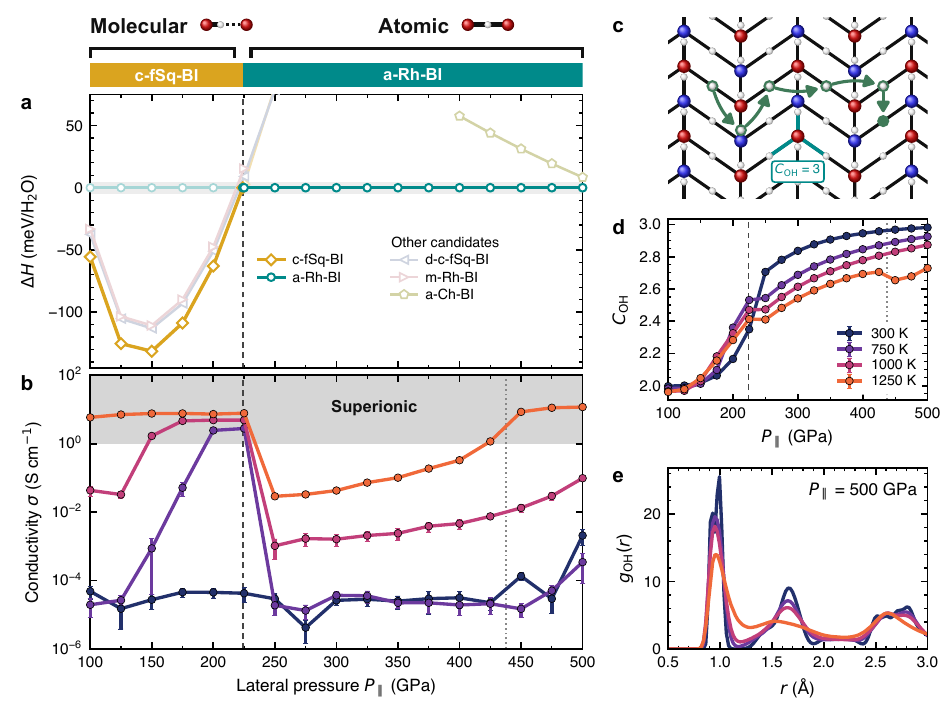}
  \caption{\textbf{Phase ordering and proton dynamics of implicitly confined water.}
  \\
  \textbf{a)} Relative enthalpy $\Delta H$ of implicitly confined $\mathrm{H_2O}$ at $T=0$~K and slit width $w=5$~\AA{} from 100 to 500~GPa, given relative to the atomic flat-rhombic phase (a-Rh-BI).
  The grey area around $\Delta H = 0$ depicts the approximate degeneracy window of $5~\mathrm{meV}/\mathrm{H_2O}$.
  The bar above indicates the most stable phase in each pressure range.
  The metastable candidates d-c-fSq-BI and m-Rh-BI are shown in the SI~\cite{supplemental}, Fig.~S6.
  \textbf{b)} Protonic conductivity $\sigma$ as a function of $P_\parallel$ at different temperatures.
  The grey region marks the superionic regime above $1~\mathrm{S\,cm^{-1}}$.
  At $300~\mathrm{K}$, hydrogen atoms remain localised throughout the entire pressure range.
  Each $(P_\parallel, T)$ point is obtained from a $1~\mathrm{ns}$ trajectory.
  \textbf{c)} Schematic of proton hopping in the a-Rh-BI ice phase.
  Green arrows trace successive proton jumps through the network.
  The highlighted bonds show the origin of $C_\text{OH} = 3$ in this phase, with each oxygen carrying two symmetrically shared in-plane hydrogens and one dangling hydrogen pointing out of the plane.
  \textbf{d)} Average \mbox{O--H} coordination number $C_{\mathrm{OH}}$ (defined in Methods) as a function of lateral pressure $P_\parallel$ at different temperatures.
  Error bars are too small to be visible.
  \textbf{e)} \mbox{O--H} radial distribution function $g_{\mathrm{OH}}(r)$ at $500~\mathrm{GPa}$ for each temperature.
  In a), b) and d), the first dashed line marks the molecular-to-atomic transition, where discrete $\mathrm{H_2O}$ units break down into an extended \mbox{H--O} bonded network.
  In b) and d), the second dotted line marks the onset of interlayer \mbox{H--O} bonding, connecting previously isolated sheets.
  }
  \label{fig:implicit-ordering-dynamics}
\end{figure*}

We map the structural response of nanoconfined water to lateral compression, imposed through the in-plane pressure $P_\parallel$, by choosing a slit consisting of two implicit uniform parallel walls separated by a width of $w=5$~\AA{}, consistent with previous work~\cite{Kapil2022,2DIce-ab-init,jiang_rich_2024}.
Since hydrogen--wall interactions are no longer negligible at the wall separations and pressures considered here, we use species-specific confining potentials for both oxygen and hydrogen, derived from an atomistically resolved Improved Lennard-Jones (ILJ) potential~\cite{vekeman_development_2023}.
Furthermore, to ensure a cost-efficient and accurate way of sampling different polymorphs at various thermodynamic conditions, we employ an iterative training procedure \cite{supplemental} to obtain an accurate MACE potential \cite{batatia2023macehigherorderequivariant} trained at the hybrid level of theory, revPBE0-D3~\cite{revPBE,PBE,PBE0,D3}.
The resulting model achieves excellent accuracy across the full pressure range considered, allowing extensive and reliable random structure searches~\cite{AIRSS} with errors of just a few meV per molecule, and hence reliable structure orderings.
Moreover, the model facilitates nanoseconds' worth of simulations at comparable accuracy for hundreds of atoms.
See SI~(\cite{supplemental}, Fig. S1) for MLP testing and error evaluation.

With these
tools at hand,
we sample the most stable structures from $10$~bar up to $550~\mathrm{GPa}$, shown schematically in Fig.~\ref{fig:structural-evolution}.
At the lowest pressures probed, between $10$~bar and $5~\mathrm{GPa}$, we find two near-degenerate phases: the monolayer flat-square and flat-rhombic ices (f-Sq/f-Rh), with f-Rh remaining stable up to $25~\mathrm{GPa}$.
In contrast to previous studies~\cite{2DIce-ab-init, Kapil2022}, which identify the hexagonal and Cairo-tiling (see Fig. S3 in SI \cite{supplemental}) structures as stable at low pressure, we instead find f-Sq/f-Rh to be favoured, owing to the employed hydrogen confining potential, 
stabilising
structures with in-plane-aligned bonds.
Considering only the oxygen lattice, the f-Sq structure has \textit{p4gm} wallpaper-group symmetry, with the water dipoles alternating antiferroelectrically along two orthogonal lattice directions.
The f-Rh structure instead has \textit{p2gg} symmetry, the dipoles again alternating along two lattice directions but now at an obtuse angle.

Proceeding to intermediate pressures, we predict the branched zigzag quasi-bilayer ice phase (b-ZZ-qBI) to be stable over the range $25$--$100~\mathrm{GPa}$.
This phase has \textit{p6mm} wallpaper symmetry, combining hexagonal translational symmetry with sixfold rotational symmetry.
It comprises two closely spaced sheets, each built from zigzag chains.
At every vertex of a chain an additional water molecule points in an alternating in-plane direction perpendicular to the propagation direction of the chain.
In this region we also sample closely related variants that differ in the branching pattern along the zigzag arms, as well as structures in which the hydrogen-bonded network is no longer molecular but atomic.
See the SI~(\cite{supplemental}, Figs. S2 to S5) for the corresponding enthalpy curves and structural snapshots of the low- and intermediate-pressure regimes.
Importantly, our findings are overall consistent with previous reports~\cite{jiang_rich_2024, lin_temperature-pressure_2023}. 

At yet higher pressures, the structural evolution gives way to a family of dense bilayer phases, whose enthalpy ordering is reported in Fig.~\ref{fig:implicit-ordering-dynamics}a.
Here we find three novel stable phases: centred flat-square bilayer ice (c-fSq-BI; $100$--$225~\mathrm{GPa}$), atomic rhombic bilayer ice (a-Rh-BI; $225$--$525~\mathrm{GPa}$), and atomic chevron bilayer ice (a-Ch-BI; $\geq525~\mathrm{GPa}$).

The c-fSq-BI structure comprises two flat square sheets in AB stacking.
The individual sheets adopt a different hydrogen-bond arrangement from the monolayer \mbox{f-Sq} phase at low pressure: within each sheet all water dipoles point in the same direction, along a square diagonal, while the adjacent sheet carries the opposite dipole, rendering the bilayer as a whole anti-ferroelectric.
Therefore, compared to the monolayer f-Sq case, the dipole flips now occur between the sheets rather than within them.
The phase has \textit{p4mm} wallpaper symmetry, in contrast to the \textit{p4gm} symmetry of f-Sq, reflecting exact mirror rather than glide symmetry.
Moreover, in c-fSq-BI the network is still molecular, with each water molecule accepting and donating two hydrogen bonds within its own sheet.

At $225~\mathrm{GPa}$ we arrive at a-Rh-BI, consisting of two flat rhombic sheets stacked in an AB scheme.
Its wallpaper symmetry is \textit{p2mg}: relative to the monolayer flat-rhombic case, the additional symmetry introduced by the AB stacking replaces a glide by a mirror.
The most striking feature of this phase, however, is the onset of the atomic regime.
Here the discrete H$_2$O units break down entirely, giving way to an extended symmetric \mbox{O--H--O} bonded network in which each oxygen takes on a coordination number of $C_\text{OH}=3$ (see Fig. \ref{fig:implicit-ordering-dynamics}c).
The same qualitative molecular-to-atomic transition occurs in bulk water, where it gives rise to ice X at roughly $120~\mathrm{GPa}$~\cite{highpressureice}, in which, however, each oxygen is symmetrically bonded to four hydrogens ($C_\text{OH}=4$).
Under confinement, the atomic network is instead three-coordinated, with no bulk counterpart.
The dipole arrangement is equally noteworthy: each sheet carries a net dipole pointing out of the plane, with the two sheet dipoles opposed, rendering this phase anti-ferroelectric.
At $500~\mathrm{GPa}$ the nearest-neighbour O--O distance is 1.84~\AA{} in-plane and 1.70~\AA{} between the sheets.

At 525 GPa, beyond the range shown in Fig.~\ref{fig:implicit-ordering-dynamics}a, a-Rh-BI gives way to a second atomic phase of the same coordination, $C_\text{OH}=3$: the a-Ch-BI phase, which has an approximate \textit{c2mm} wallpaper symmetry.
This phase marks the next characteristic transition, from structurally independent sheets to an interpenetrating hydrogen network that connects them.
At $525~\mathrm{GPa}$ its nearest-neighbour O--O distance is 1.84~\AA{} in-plane and 1.76~\AA{} between the sheets.

To further strengthen our predictions, we have probed the influence of nuclear quantum effects on the relative stability of the closely competing high pressure structures.
We have computed the harmonic zero point energy (ZPE) of the a-Rh-BI and a-Ch-BI phases at 500~GPa, where the competing phases lie closest in enthalpy.
The a-Rh-BI phase lies higher in ZPE than a-Ch-BI by $\Delta E_\text{ZPE} \sim24~\text{meV/water}$ (see SI~\cite{supplemental}, Fig. S7), which shifts the a-Rh-BI~$\to$~a-Ch-BI transition to lower pressure by up to $\sim50~\text{GPa}$.
This difference is typical of competing high-pressure ice phases, whose ZPEs usually differ by less than $\sim50~\text{meV/water}$~\cite{highpressureice}.
Corrections of this size are expected to shift the transition pressures by at most tens of GPa, leaving the sequence of transitions reported here intact.

\subsection*{High-pressure superionicity in confined water}

High-pressure environments, such as those encountered in shock-compression experiments or planetary interiors, are often accompanied by high temperatures~\cite{Militzer_2008, InteriorOfGiantPlanets, millot_nanosecond_2019}.
Moreover, temperature itself is known to drive rich physics in both bulk and nanoconfined water~\cite{jiang_rich_2024, Kapil2022, iceX-hopping}.
We build upon the identified ice phases and thermodynamically sample a vast temperature range to assess the effect of thermal fluctuations on both the atomic and electronic structure.
Explicitly, at each pressure point $P_\parallel$ in the range $100$--$500~\mathrm{GPa}$ we simulate target temperatures between 300 and 1250\,K.

We observe a pronounced superionic character at high temperatures at both ends of the thermodynamically sampled pressure regime.
As visible in Fig.~\ref{fig:implicit-ordering-dynamics}b, we find the phase diagram to consist of two separated superionic regions ($T = 1250$~K), indicated by the ionic conductivity $\sigma$ exceeding $\sim1~\text{S}~\text{cm}^{-1}$, the stricter of the two criteria commonly used ~\cite{Kapil2022, iceX-hopping}.
The first transition, from a superionic to a non-superionic regime, is driven by the c-fSq-BI~$\to$~a-Rh-BI structural transition in Fig.~\ref{fig:implicit-ordering-dynamics}a.
In line with the drop in conductivity, we compute lower proton diffusion barriers in the molecular, square hydrogen-bonded c-fSq-BI phase. 
A particularly pronounced difference appears in the angular free-energy profile $F(\phi)$ of the O--H bonds, whose barriers are significantly higher in the a-Rh-BI phase (see SI~\cite{supplemental}, Figs.~S8 and S9).
The molecular superionic dynamics of the c-fSq-BI phase are depicted in SI Video~1.

The second crossover, where the conductivity rises by roughly an order of magnitude, is likewise accompanied by a structural transition, from a-Rh-BI to a-Ch-BI.
Unlike the first transition at 225 GPa, which follows from the 0 K enthalpy ordering, this one is thermally induced and occurs at ${\sim}435~\mathrm{GPa}$ at $1250~\mathrm{K}$, about 90~GPa below the static transition at 525~GPa.
Since a-Rh-BI and a-Ch-BI share closely related oxygen lattices, this transition is essentially a thermally driven reorganisation of the hydrogen network, in which a subpopulation of hydrogens reorients towards the opposing sheet and forms interpenetrating connections between the sheets.
This interlayer-connecting character -- illustrated in SI Video~2 -- opens hydrogen-transport pathways between the two sheets and drives substantially higher hydrogen diffusion throughout the lattice, thereby increasing the ionic conductivity.

Furthermore, both transitions are accompanied by distinct changes in the oxygen-hydrogen coordination number $C_\text{OH}$ (see Fig.~\ref{fig:implicit-ordering-dynamics}d), reflecting the molecular-to-atomic transition and the onset of interlayer hydrogen reorganisation, respectively.
Up to the first transition, $C_\text{OH}$ exhibits almost no temperature dependence: in the molecular phases, each hydrogen is tightly bound to one oxygen, an arrangement too well-defined to be disrupted by thermal motion.
Beyond the first transition, in the atomic phases, lattice fluctuations within the extended O$-$H network allow for hydrogen asymmetry between two oxygen atoms, driving an increasing deviation of $C_\text{OH}$ from its ideal, zero-temperature value $C_\text{OH} = 3$.

As for the nature of the superionic phases, different pressure regimes are found to induce different hydrogen diffusion mechanisms.
At lower pressures, we find a molecular superionic phase with no bulk counterpart, characterised by the Grotthuss mechanism as reported in past molecular nanoconfined water studies \cite{coles2025nanoconfinedsuperionicwatermolecular}. 
Here the free energy profile $F(\delta)$ of the proton transfer coordinate $\delta = r_{\mathrm{O}{-}\mathrm{H}}-r_{\mathrm{O}\cdots\mathrm{H}}$ exhibits two wells (see Fig. S8 in SI \cite{supplemental}).
Proton conduction therefore involves two elements: transfer along an asymmetric hydrogen bond $\mathrm{O{-}H\cdots O}$ to a neighbouring oxygen, and rotation of an O--H bond about its oxygen, which brings a proton into position for a further transfer.
This behaviour is illustrated in the SI Video~3.
One difference from monolayer low pressure confined superionic phases \cite{coles2025nanoconfinedsuperionicwatermolecular} is the high defect concentration of up to ${\sim}20\%$, which precludes tracking the diffusion of individual defects across the lattice.

At higher pressures the character of the superionic phase changes substantially, indicated by the high coordination number approaching $C_\text{OH} \to 3$: the molecular superionic phase has given way to an atomic one.
Here we observe a single well in the free-energy profile (see SI \cite{supplemental}, Fig. S10), indicating that protons are predominantly located in symmetric bonds $\mathrm{O{-}H{-}O}$, midway between neighbouring oxygen atoms.
While thermal fluctuations still give rise to proton displacements about this central position, the proton-transfer mechanism is now dominated by angular hops of the hydrogens about the oxygen atoms rather than by translations between distinct bonding sites.
An illustration of this mechanism is provided in SI Video~4, while the resulting hydrogen diffusion pattern is shown schematically in Fig.~\ref{fig:implicit-ordering-dynamics}c.
Note that a similar proton diffusion mechanism is found in bulk in superionic ice~X~\cite{iceX-hopping}.
Similar to bulk superionic phases \cite{coles2025nanoconfinedsuperionicwatermolecular} at high pressures and temperatures, we observe in both confined superionic regions a transition from a solid hydrogen structure to a liquid-like one, which can be inferred from the peak broadening and overlap in the $g_\text{OH}(r)$ radial distribution function (RDF).
This is shown in Fig.~\ref{fig:implicit-ordering-dynamics}e for the 500~GPa, 1250~K state point; additional O--O and O--H RDFs spanning 100--500~GPa are provided in Fig. S11 of the SI~\cite{supplemental}.

Notably, at high pressures the superionic character need not be thermally driven.
At $T=0~\mathrm{K}$ and 500~GPa, the underlying a-Rh-BI ice phase exhibits a total hydrogen-projected ZPE of $E^{H}_\text{ZPE}\sim1~\mathrm{eV/water}$, with hydrogen vibrational frequencies of up to $4500~\mathrm{cm}^{-1}$ (0.55~eV), while the oxygen-projected ZPE is negligible.
Since the hydrogen ZPE is of the same order as the concerted H-diffusion barriers predicted for high-pressure bulk ices ($\sim0.7~\mathrm{eV}$)~\cite{highpressureice}, zero-point motion is likely to further enhance proton mobility.
It
may even melt the hydrogen sublattice at low temperatures, yielding a ``cold'' superionic phase in which the oxygen structure remains intact.

\subsection*{Superionicity persists under explicit diamond confinement}

\begin{figure*}[t]
  \centering
  \includegraphics{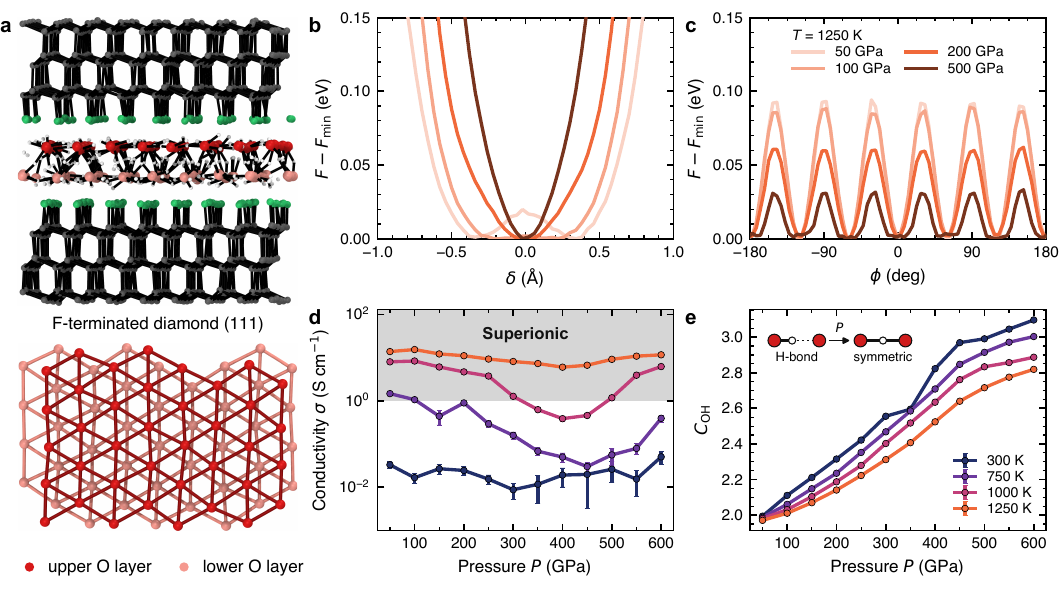}
\caption{\textbf{Structure and proton dynamics of water in a diamond anvil cell.}
\\
\textbf{a)} Top: side view of the simulation cell ($\sim 2000$ atoms, including 128 water molecules), with a water bilayer confined between two F-terminated diamond (111) slabs. For clarity, only the carbon atoms nearest to the water are shown.
Carbon atoms are shown in grey, fluorine in green, hydrogen in white, and oxygen in dark and light red for the upper and lower water layer, respectively.
Bottom: top view of the oxygen atoms of the bilayer.
Lines connect nearest-neighbour O atoms within each layer, showing two stacked triangular O lattices, each offset into the hollow sites of the other.
\textbf{b)} Free energy profile $F - F_\mathrm{min}$ along the proton transfer coordinate $\delta = r_{\mathrm{O-H}} - r_{\mathrm{O\cdots H}}$ for protons in \mbox{O--H$\cdots$O} configurations at $1250~\mathrm{K}$ and different pressures.
\textbf{c)} Free energy profile of the same protons along the \mbox{O--H} bond orientation, parametrised by the azimuthal angle $\phi$ within the confining ($xy$) plane.
In b) and c), each profile is shifted to its own minimum.
\textbf{d)} Protonic conductivity $\sigma$ as a function of isotropic pressure $P$ at different temperatures, where the shaded region indicates the superionic regime above $1~\mathrm{S\,cm^{-1}}$.
\textbf{e)} Average \mbox{O--H} coordination number $C_{\mathrm{OH}}$ (defined in Methods).
The inset illustrates the pressure-driven change from \mbox{O--H$\cdots$O} hydrogen bonds to symmetric \mbox{O--H--O} bonds.
Each $(P, T)$ point is obtained from a trajectory of ${\sim}200~\mathrm{ps}$.
}
  \label{fig:explicit-dynamics}
\end{figure*}

Explicit confining materials have been employed in previous studies~\cite{Kapil2022, metallicHydrogen}.
However, these setups either treat the confining material as rigid, which is unphysical, or rely on materials such as graphene, restricting the accessible range to a few gigapascals.
This prevents pushing to higher pressures and, more importantly, connecting the low-, intermediate- and high-pressure regimes reliably and consistently.
Here we introduce a setup that overcomes both limitations, and systematically maps the response of nanoconfined water to compression, this time in the realistic environment of a diamond anvil cell model.
As depicted in Fig.~\ref{fig:explicit-dynamics}a, our setup consists of two diamond slabs cut along the (111) surface.
Since cleaving diamond leaves reactive dangling bonds -- passivated in experimental setups -- we terminate them with fluorine, providing a chemically inert slit of tunable width.
Other terminations, such as hydrogen, oxygen or chlorine, are possible~\cite{FterminatedD}, giving rise to many variants of the diamond anvil cell.

To probe the dynamic stability of the molten hydrogen lattice in this environment, we first filled the slit with water, relaxed the cell and ran finite temperature \textit{ab initio} MD simulations at the PBE level.
Not only does the structure remain stable, it also retains signs of superionic character (see SI Video~5 and SI~\cite{supplemental}, Fig.~S12, for computational details).
Such \textit{ab initio} simulations are, however, restricted to small cells and short trajectories, whereas a realistic diamond-anvil geometry requires thousands of atoms for only a few hundred water molecules.

We therefore trained EDDPs~\cite{EDDPlammps, EDDPrandomstructure} on PBE reference data and ran nanosecond-scale simulations on systems of up to 4000 atoms within the diamond-anvil cell, from ambient temperature up to 1250~K (see SI Video~6 and SI~\cite{supplemental} for computational details).

With these potentials, we sample a wide region of the phase diagram, from 50 to 600~GPa and from~300~to~1250~K.
Here we recover the two distinct superionic regimes found under implicit confinement, a molecular one at low pressures and an atomic one at high pressures.
At $T = 1250~\mathrm{K}$, both regimes are marked by a strong superionicity ($\sigma \gtrsim 10^1~\mathrm{S\,cm^{-1}}$, Fig.~\ref{fig:explicit-dynamics}d), while the coordination number rises from $C_\mathrm{OH} = 2$ at 50~GPa towards $C_\mathrm{OH} = 3$ at 600~GPa (Fig.~\ref{fig:explicit-dynamics}e).
Furthermore, despite the electronically and thermodynamically different confinement environment, the transport mechanisms of both regimes are preserved.
The free energy along the proton transfer coordinate evolves from a double well at low pressure into a single well (Fig.~\ref{fig:explicit-dynamics}b), reflecting the change from the molecular hopping regime to the symmetrised one, in which the hydrogen sits midway between two oxygens.
At the same time, the rotational free-energy barriers decrease with increasing pressure (Fig.~\ref{fig:explicit-dynamics}c), 
facilitating hydrogen bond rotations,
allowing the hydrogens to diffuse easier through angular hops about the host oxygen, which drives the recovery of the conductivity in the atomic regime.
The hallmark conductivity dip separating the two regimes is likewise preserved, with the conductivity falling below the superionic threshold at 1000~K and essentially vanishing at 750~K (Fig.~\ref{fig:explicit-dynamics}d).

Interestingly, the oxygen lattice is triangular at this confinement width, matching the arrangement of the terminating fluorines on the (111) surface, with each oxygen sitting on average within a fluorine triangle.
The width is set by the amount of water packed into the slit, which here forms a bilayer.
As a result, the structure of the oxygen lattice does not change with pressure, and only the distances $d_\mathrm{OO}$ shrink.
At 300~K, the nearest-neighbour in-plane distance $d_\mathrm{OO}$ decreases from 2.28~\AA{} at 50~GPa to 2.03~\AA{} at 600~GPa, and the out-of-plane one from 2.45~\AA{} to 1.98~\AA{}.
Both $d_\mathrm{OO}$ and the volume per formula unit track the bulk ice X values closely over the overlapping pressure range (SI~\cite{supplemental}, Figs.~S14 and S16), confirming that the explicit setup samples pressures comparable to bulk studies of high-pressure ice.
Even without any oxygen-lattice transition, we recover the same structural and dynamical hallmarks: the molecular-to-atomic transition with its characteristic evolution of the free-energy barriers, and the superionic transitions with the conductivity minimum between them.
This points to a local origin of these hallmarks rather than one set by the global crystal structure.

\subsection*{Electronic structure under confinement}

\begin{figure*}[t]
  \centering
  \includegraphics{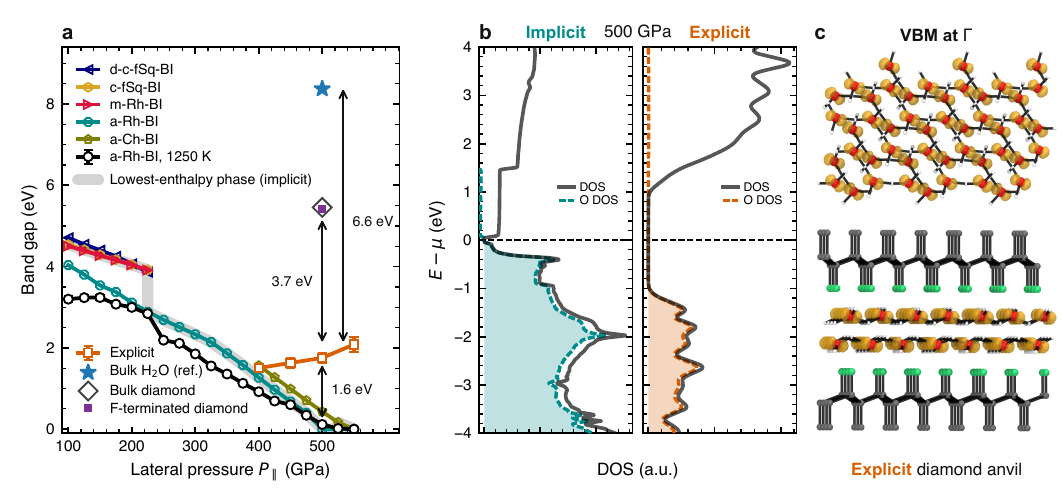}
  \caption{\textbf{High-pressure electronic structure of water under implicit and explicit confinement.}
  \\
  \textbf{a)} Band gap at the PBE level for the implicitly confined ice phases shown in Fig.~\ref{fig:structural-evolution} and for a finite-temperature a-Rh-BI phase.
  The grey band follows the lowest-enthalpy phase under implicit confinement at T = 0~K.
  Orange squares show water in the explicit diamond anvil cell at finite temperature, compared with bulk $\mathrm{H_2O}$~\cite{highpressureice}, bulk diamond and F-terminated diamond at $500~\mathrm{GPa}$.
  For the finite-temperature values ($1250~\mathrm{K}$), the gap at each pressure is averaged over statistically independent frames.
  Arrows mark the gap differences relative to the explicit system at $500~\mathrm{GPa}$. For the explicit system, the horizontal axis gives the isotropic pressure $P$ of the full cell rather than $P_\parallel$.
  All gaps are computed with PBE, which underestimates band gaps.
  At the HSE06 level, gap closure of a-Rh-BI shifts from 500 to 725~GPa (see SI~\cite{supplemental}, Fig.~S17).
  \textbf{b)} Total and oxygen-projected electronic DOS at $500~\mathrm{GPa}$ around the chemical potential $\mu$, under implicit (left, a-Rh-BI) and explicit (right) confinement.
  In both cases, the top of the valence band is dominated by oxygen 2$p$ lone-pair states.
  \textbf{c)} Valence band maximum (VBM) wave function at the $\Gamma$~point in the explicit diamond anvil cell.
  Top: top view of the water layer, with carbon and fluorine omitted for clarity.
  Bottom: side view of the full cell.
  The VBM is localised on the oxygen atoms and retains its lone-pair character under explicit confinement.
  Carbon atoms are shown in grey, fluorine in green, oxygen in red and hydrogen in white.
  }
  \label{fig:electronic}
\end{figure*}

Having established how confinement reshapes the atomic structure and dynamics, we now turn to its effect on the electronic structure.
That confinement and dimensionality can substantially restructure electronic states is well established: in van der Waals layered materials, for instance, reducing the number of layers strengthens quantum confinement and is known to open the band gap, as interlayer coupling weakens~\cite{quantum_confinement}.
Here we consider a different regime: a hydrogen-bonded network under extreme lateral pressures, where confinement acts not through interlayer coupling but through the bonding topology itself.
In this high-pressure, hydrogen-rich setting, we find that the band gap is substantially reduced relative to bulk water.
At the same time, the top of the valence band retains, up to hundreds of gigapascals, the oxygen 2p lone-pair character familiar from ambient water and ice.
Whether the gap closes, however, depends on the confining material: between implicit walls it closes, whereas between explicit diamond walls it remains open.

Under implicit confinement, the band gap of all stable high-pressure structures follows the same downward trend towards zero with increasing pressure (Fig.~\ref{fig:electronic}a).
At the GGA level (PBE~\cite{PBE}), a-Rh-BI closes at $500~\text{GPa}$ and a-Ch-BI shortly after at $550~\text{GPa}$.
For the superionic a-Rh-BI phase at 1250~K, gap closure falls between these two values, at $\sim$525~GPa, showing that temperature has little effect on the gap-closure pressure.
At the hybrid (HSE06~\cite{HSE06}) level, gap closure for a-Rh-BI shifts upward to 725~GPa (SI~\cite{supplemental}, Fig.~S17).
Closure occurs at the $\Gamma$~point, between purely oxygen-derived valence states and purely hydrogenic conduction states (Fig.~\ref{fig:electronic}b, left).
These hydrogenic conduction states, however, delocalise out of the confinement plane into the vacuum region beyond the implicit walls (Fig.~S19).
In a real confining material, that region is occupied by the material's own atoms and electronic states.

To test the implicit-wall picture against a realistic environment, we turn to the explicit F-terminated diamond anvil cell introduced above.
We fully relaxed it under an isotropic pressure of 500~GPa at the PBE level (SI~\cite{supplemental}, Figs.~S12 and S13).
Here, the gap does not close, instead increasing slightly with pressure.
The relaxed structure ($T=0$~K) retains a gap of $\sim$2.1~eV (Fig.~S13).
At finite temperature the gap remains open across all sampled pressures, lying 1.6~eV above the implicit-wall value at 500~GPa (Fig.~\ref{fig:electronic}a).
The valence manifold behaves as under implicit confinement: its top is formed by pure oxygen states (Fig.~\ref{fig:electronic}b), and the valence band maximum (VBM) wave function at $\Gamma$ remains localised on the oxygen atoms with 2p lone-pair character (Fig.~\ref{fig:electronic}c).
The conduction manifold, in contrast, changes character.
The hydrogenic states are pushed up substantially, and the conduction band minimum is now formed by delocalised carbon states of the diamond (Fig.~S13).

This reordering reflects the band alignment between the two materials, together with the confinement of the hydrogenic states, which can no longer delocalise beyond the confinement plane.
The oxygen valence states lie $\sim$3~eV above the carbon valence band maximum, inside the band gap of the diamond, directly below its carbon conduction states (Fig.~S13).
The gap is therefore set across the interface, between the oxygen lone-pair states of the water and the carbon conduction states of the diamond.
At 500 GPa the gap of the combined system lies 3.7 eV below the gap of diamond and 6.6 eV below that of bulk water (Fig.~\ref{fig:electronic}a).

The oxygen lone-pair character of the valence band is thus common to both setups, as seen in the VBM wave function at $\Gamma$ under explicit confinement (Fig.~\ref{fig:electronic}c) and in the valence-band maximally-localised Wannier functions (MLWFs) under implicit confinement (Fig.~S18).
The MLWFs of the a-Rh-BI phase are centred on the oxygen atoms.
Their 2p lobes point out of plane, away from the covalent O--H bonds, retaining the lone-pair character of low-pressure ice~\cite{hexdensity}.
In the three-coordinated atomic network ($C_\text{OH}=3$) identified above, the confinement geometry leaves no bonding partners along these directions, which prevents re-hybridisation and preserves the lone pairs, and compression instead raises their band edge in energy (see SI, Fig. S19).
This is unlike high-pressure bulk ice, where hydrogen-bond symmetrisation absorbs the lone pairs into sp$^3$-like bonding orbitals, as found for example in ice~X~\cite{icexwannier}.

Confinement thus acts differently on the valence and conduction manifolds.
The confined water sets the valence band, whose oxygen lone pairs persist under compression.
The confining material sets the conduction band, and with it whether the gap closes.
The band gap of nanoconfined water is therefore a property not of the water alone but of the water--wall interface.

\section{Conclusions}
\label{sec:conclusions}
We have shown that nanoscale confinement fundamentally restructures the structural, dynamical and electronic evolution of water under extreme pressure.
Mapping confined water within one consistent framework from ambient conditions to hundreds of gigapascals, we report three novel bilayer phases with no known bulk analogue.
We also find a bulk high-pressure hallmark, the molecular-to-atomic transition.
Under confinement, however, it yields a three-coordinated atomic network ($C_\text{OH}=3$, versus 4 in ice X).
Furthermore, superionicity likewise occurs in both a molecular and an atomic regime, separated by a conductivity minimum at the molecular-to-atomic crossover: protons move by the Grotthuss mechanism in the former and by angular hops around the oxygens in the latter.
The molecular regime has no bulk counterpart.
Bulk superionic ice changes character as its hydrogen bonds symmetrise~\cite{iceX-hopping} but never remains molecular~\cite{coles2025nanoconfinedsuperionicwatermolecular}.
Whereas bulk water reaches atomic superionicity in ice X at roughly 120~GPa and 1800~K~\cite{iceX-hopping}, confined water remains molecular and superionic up to 225~GPa and becomes an atomic superionic
state
only above $\sim$435~GPa, already at 1250~K.

These regimes persist from implicit to explicit confinement, despite the different interatomic potential and reference functional.
Between explicit, fully flexible diamond walls, both regimes and a conductivity minimum between them survive even without any oxygen-lattice transition, pointing to a local origin rather than one set by the global crystal structure.
This setup resembles the geometry of diamond-anvil-cell experiments on superionic bulk ice~\cite{natureDAC, prlDAC}.

Electronically, in bulk ice, compression absorbs the oxygen lone pairs into the symmetric bonding network~\cite{icexwannier}.
Under confinement, it instead preserves them and raises their band edge in energy.
Whether the gap closes, however, is controlled by the confining material.
Implicit walls close the gap at $\sim$500~GPa at the PBE level, whereas explicit diamond walls keep it open at about 1.8~eV, still 6.6~eV below that of bulk water~\cite{highpressureice} and 3.7~eV below that of diamond at the same pressure
and computational setup.
The difference between implicit and explicit confinement arises from band alignment, which places the carbon conduction states below the hydrogenic ones.
The confining material thus acts as an active ingredient rather than a passive wall, setting the electronic response of the confined water.
Varying the surface termination of the confining material~\cite{FterminatedD} offers a direct route to test and exploit this control.

Overall, we establish confinement as a thermodynamic axis alongside pressure and temperature, one that selects phases rather than merely shifting their boundaries.
It decouples superionicity from hydrogen-bond symmetrisation, and the resulting conductivity minimum between the two regimes provides a
measurable transport property
that requires no structural resolution.
For simulation practice, implicit walls capture structure and proton dynamics but not the electronic response, so electronic properties of confined matter require explicit walls.
Finally, the underlying mechanism is geometric: confinement removes bonding partners and thereby preserves the oxygen lone pairs.
It should therefore carry over to other hydrogen-rich systems, opening a general frontier for two-dimensional materials at high pressure.

\section{Methods}
\label{sec:computational-methods}

Here we give a concise summary of the employed methods and parameters.
For a more extensive discussion, additional figures and data, consult the SI~\cite{supplemental}.

\textbf{Obtaining the confining potential.}
We derive the implicit confining potential from an ILJ potential in Ref.~\cite{vekeman_development_2023}, fitted at the GGA level of theory (B97-D) to circumcoronene and benchmarked against Coupled Cluster theory.
Its form is species-specific and additionally includes an induction contribution,
\begin{equation}
    U= \sum_{ij}V_{ij}(R_{ij}) - \frac{1}{2}\sum_{i\in\text{graphene}} \alpha_i\bigg(\sum_{j\in\text{water}}\frac{q_j}{R_{ij}}\bigg)^2 ,
\end{equation}
where $V_{ij}(R_{ij})$ is the ILJ contribution between atom $i\in\text{graphene}$ and atom $j\in\text{water}$, and the second term accounts for the induced polarisation of the graphene sheet by the water charges $q_j$.
We omit this induction term throughout, as it does not significantly alter the repulsive branch of the potential, which is the branch that dominates the sampled structures in the narrow slits employed here.
This omission allows the remaining pairwise sum to be cleanly separated into independent oxygen and hydrogen contributions.
With this, we evaluate the resulting potential explicitly for a single water molecule placed above a finite sheet of carbon atoms
\begin{equation}
    V(z) = \sum_j \big[V_{H_1}(R_{jH_1}) + V_{H_2}(R_{jH_2}) + V_{O}(R_{jO})\big] ,
\end{equation}
where the sum runs over all carbon sites $j$ in the sheet and $z$ denotes the perpendicular distance of the water molecule from the sheet and $R_{jH_1}$ is the distance from one hydrogen atom to the $j$-th carbon.
We converge this sum with respect to sheet size until $V(z)$ no longer changes appreciably.
Having fixed the converged sheet, we then vary the perpendicular coordinate $z$ of the water molecule to obtain and tabulate $V_O(z)$ and $V_H(z)$.

\textbf{Modelling the pressure control.}
Analysing pressure-driven evolution in 2D systems requires a consistent definition of the controlled pressure quantity, here the lateral pressure $P_\parallel$.
Since atomic motion is restricted in the confinement direction, the corresponding component of the pressure tensor cannot be controlled meaningfully, as it represents the independent response of the system to the confining material.
Therefore, the perpendicular pressure \(P_\perp = \sigma_{zz}\) is decoupled from the barostat throughout.
Here, \(\sigma = \sigma' L_z/w\) is the rescaled stress tensor, where \(\sigma'\) is the computed stress tensor, \(L_z\) is the full simulation-cell length along the confinement direction, and \(w\) is the confinement width.
Similarly, the lateral pressure is defined as
\begin{equation}
P_\parallel = \frac{\sigma_{xx}+\sigma_{yy}}{2},
\end{equation}
where the \(x\)- and \(y\)-components are coupled and fluctuate jointly during the simulations.
We take the confinement width \(w\) to be equal to the spacing between the implicit walls, following a standard convention used in simulations of nanoconfined water~\cite{2DIce-ab-init,Kapil2022,jiang_rich_2024}.
This choice, however, represents an upper bound to the accessible layer height, since the steep wall potential reduces the effective phase space available to the atoms.
The exact accessible width depends on temperature, pressure, and the shape of the confining potential and is therefore difficult to be determined uniquely.

Changing this definition does not alter the overall shape of the pressure dependent curves, but can rescale the corresponding pressure values.
In the high-pressure structures studied here, the observed atomic positions span at least half of the wall-to-wall width, providing a conservative lower bound \(w_\mathrm{eff} \geq w/2\).
Thus, even under the most conservative effective-width convention, the reported transition pressures would increase by at most a factor of two.

\textbf{Ice enthalpy ordering.}
To order the resulting implicitly confined high-pressure phases, we compute the relative enthalpy of the confined system at $T=0$~K, defined per water molecule as
\begin{equation}
    H = E_\mathrm{water} + E_\mathrm{walls} + P_\parallel \times \, w \times \, A ,
\end{equation}
where $E_\mathrm{water}$ is the total energy per water molecule, $E_\mathrm{walls}$ is the energy of the two confining walls per water molecule, $w = 5$~\AA{} is the confinement width, and $A$ is the lateral area per water molecule. 

\textbf{Development of the water MLP.}
We develop a MACE MLP for water at high pressures, with the dataset from Ref.~\cite{Kapil2022} serving as the seed for the iterative learning procedure.
All force and energy computations were performed at the revPBE0-D3 level of theory.
Furthermore, we employ the MACE-OMAT \cite{MACEMP0} foundation model to sample finite-temperature structures at moderately high pressures, with the reach of the model extending into the low hundreds of GPa, and add these structures to the dataset.
We then push the obtained model to progressively higher pressures in finite-temperature MD simulations until its breaking point is reached, relabel frames shortly before the onset of instabilities, and add them to the dataset.
This procedure is repeated until a model that is stable and accurate up to the desired pressure regime of approximately $500~\mathrm{GPa}$ is obtained.
At all stages of the data sampling procedure, we employ the implicit walls described above. 
Furthermore, for inference, we use \textit{symmetrix}~\cite{witt2025symmetrix, MACEMP0}, a CPU implementation of MACE that enables massive parallelization for extensive random structure searching and MD sampling.

\textbf{EDDPs for the diamond anvil cell.}
For the explicit diamond anvil cell, we train ephemeral data-derived potentials (EDDPs)~\cite{EDDPrandomstructure, EDDPlammps} on ${\sim}10{,}000$ random structures of the diamond--water system, labelled at the PBE level~\cite{PBE} using CASTEP~\cite{castep}.
These reference calculations use ultrasoft pseudopotentials, a plane-wave cutoff of $340~\mathrm{eV}$, and a Monkhorst--Pack $k$-point spacing of \mbox{$2\pi \times 0.1~\text{\AA}^{-1}$}.
The resulting energy errors are 12.9~meV/atom on the training set and 22.6 and 23.3~meV/atom on the validation and test sets, with Spearman rank correlations exceeding 0.999 across all splits.
All EDDP simulations are performed in the $NPT$ ensemble under isotropic pressure using the \mbox{\textsc{ramble}} molecular dynamics code of the EDDP package, with all atoms free to move.
The equations of motion are integrated with the standard velocity-Verlet algorithm at a timestep of 0.5~fs.
The temperature is controlled by Berendsen-type weak-coupling velocity rescaling (coupling time $\sim0.25$~ps), and the pressure by a fully flexible, damped variable-cell scheme (cell relaxation within~$\sim 0.5$~ps). 

\textbf{Random structure searches.}
We search for implicitly confined water ice structure candidates using the AIRSS software \cite{AIRSS}.
To this end, we randomly place between 2 and 24 water molecules into a constrained simulation cell with variable \(x\)- and \(y\)-dimensions and a fixed confinement width of \(5~\text{\AA}\) along \(z\).
Additionally, a $10$-\AA{} vacuum slab is added above the structure along the confinement direction, to prevent spurious interactions between periodic images in the subsequent electronic structure calculations.
Moreover, to generate more sensible initial configurations for relaxation and speed up the searches, up to four symmetry operations are employed during the structure generation process.
Subsequently, the generated configurations are relaxed and ranked using the MACE MLP and the implicit repulsive walls.
At each pressure point, approximately 10,000 candidate structures are sampled and evaluated.
Finally, the reliability of the resulting enthalpy ordering is assessed through revPBE0-D3 calculations on the lowest-enthalpy and otherwise most promising candidates.

\textbf{Computational details.}
Besides generating and validating the MLP training data, DFT calculations are also employed for the electronic-structure analysis.
For the former we use CP2K \cite{cp2k} with computational settings following Ref.~\cite{Schran2020}.
Specifically, we employ the revPBE0-D3 exchange--correlation functional, TZV2P \cite{cp2kBasisSet} basis sets for the valence states, and a plane-wave density representation with a cutoff of 400~Ry.
Exact exchange is evaluated using a truncated Coulomb interaction with a cutoff radius of \(6~\text{\AA}\).
Moreover, we employ Goedecker--Teter--Hutter pseudopotentials~\cite{GTH}.
We validate these against all-electron calculations and find good agreement up to high pressures (see SI~\cite{supplemental}).

In the main text, band structures are calculated with the all-electron code FHI-aims \cite{AIMS} using the \textit{intermediate} basis set and the PBE functional \cite{PBE}, enabling direct comparison with the high-pressure bulk-ice literature.
Brillouin-zone integrations are performed using dense two-dimensional $k$-point meshes of up to $50\times50\times1$. 
Ionic conductivities are estimated from the hydrogen diffusion coefficient using the Nernst--Einstein relation,
\begin{equation}
\sigma_\mathrm{NE}=\frac{n_\mathrm{H}e^2D_\mathrm{H}}{k_\mathrm{B}T},
\end{equation}
where \(n_\mathrm{H}\) is the hydrogen number density and \(D_\mathrm{H}\) the hydrogen diffusion coefficient. 

The oxygen--hydrogen coordination number is computed using a sharp switching function
\begin{equation}
C_{\mathrm{OH}}=\left\langle \frac{1}{N_\mathrm{O}}\sum_{i\in \mathrm{O}}\sum_{j\in \mathrm{H}}
\frac{1-\left(r_{ij}/r_0\right)^n}
{1-\left(r_{ij}/r_0\right)^m}
\right\rangle ,
\end{equation}
with \(r_0=1.10~\text{\AA}\), \(n=55\), and \(m=80\).
Contributions with \(r_{ij}>1.5~\text{\AA}\) are set to zero.
For the coordination number computations in the explicit confinement setup we employ a slightly elongated cutoff \(r_0=1.125~\text{\AA}\), due to the GGA functional used to create the training data for the EDDPs, as GGA functionals tend to elongate bonds in water compared to hybrid functionals. 

Implicitly confined molecular dynamics simulations are performed in LAMMPS \cite{LAMMPS} using deuterium instead of hydrogen, allowing the use of a timestep of \(1~\mathrm{fs}\).
At a given pressure point $P_\parallel$ in the range $100 -500~\mathrm{GPa}$ (Fig.~\ref{fig:implicit-ordering-dynamics}b, d) we choose the most stable structure in Fig.~\ref{fig:implicit-ordering-dynamics}a and perform a finite temperature MD simulation within the \(NP_{\parallel}T\)  ensemble at target temperatures between $300-1250~\text{K}$.
Temperature and lateral pressure are controlled using a Nos\'e--Hoover thermostat and barostat with damping constants of \(0.1~\mathrm{ps}\) and \(0.5~\mathrm{ps}\), respectively.
The \(x\)- and \(y\)-components of the pressure tensor are coupled and the lattice constant in the confinement direction ($\hat{z}$) is fixed.

Maximally localised Wannier functions (SI~\cite{supplemental}, Figs.~S18 and S19) are constructed using Wannier90 \cite{Wannier90} from PBE wavefunctions generated with SIESTA \cite{SIESTA, SIESTA2}.
We compute 40 bands by employing a TZDP~\cite{SiestaBasisSet} basis, a mesh cutoff of 400~Ry and a \(5\times5\times1\) \(k\)-point grid.
The subsequent wannierisation is performed using two Wannier functions per water molecule initialised from oxygen \(s\) and \(p_z\) projections.
We employ a disentanglement window of \([-10,10]\)~eV and a frozen window of \([-2.5,2.5]\)~eV around $E_F$.


\section*{Data availability}
The data supporting the findings of this study, together with all input files and the scripts used to generate the figures, are available at \href{https://github.com/fast-group-cam/high-p-phase-transitions}{https://github.com/fast-group-cam/high-p-phase-transitions}.

\section*{Code availability}
All codes used in this work are open source and referenced in the Methods.

\section*{Acknowledgments}
We dedicate this work to the memory of Pavel Jungwirth (1966–2026).
The ideas developed here grew directly out of discussions with him at the 2024 Gordon Research Conference on Water and Aqueous Solutions, where his enthusiasm for pushing water to its electronic limits first drew our attention to the problem.
His curiosity and generosity will be sorely missed by our community.
We thank Peter Cooke for help with the random structure searches.

I.G.D. acknowledges support from the Engineering and Physical Sciences Research Council (EPSRC) through a UKRI studentship (project reference 2928609).
A.M. thanks the European Union for support under the ``n-AQUA'' European Research Council project (Grant No. 101071937).
C.S. acknowledges support from the Isaac Newton Trust (G122390), the UKRI Critical Mass grant (EP/V062654/1) and the Royal Society (RGS/R2/242614).
We are grateful for computational support and resources from the UK national high-performance computing service, Advanced Research Computing High End Resource (ARCHER2).
Access for ARCHER2 was obtained via the UK Car--Parrinello consortium, funded by EPSRC grant reference EP/X035891/1.
The authors acknowledge the use of resources provided by the Isambard-AI National AI Research Resource (AIRR).
Isambard-AI is operated by the University of Bristol and is funded by the UK Government's Department for Science, Innovation and Technology (DSIT) via UK Research and Innovation; and the Science and Technology Facilities Council [\texttt{ST/AIRR/I-A-I/1023}].
Access to CSD3 was obtained through a University of Cambridge EPSRC Core Equipment Award EP/X034712/1.

\section*{Author contributions}
I.G.D.\ led the project, developed the computational methodology, performed all simulations and analysis, and wrote the manuscript.
C.S.\ conceived and supervised the project.
C.J.P.\ developed the EDDPs.
All authors discussed the results and edited the manuscript.

\section*{Competing interests}
The authors declare no competing interests.

\nocite{omat-dataset,phonopy-phono3py-JPCM,phonopy-phono3py-JPSJ,reinhardt_thermodynamics_2022_doo_ice_x,osmond_ultrahigh_2026_volume_blyp}

\bibliography{references}

\end{document}


\def\mytitle{From molecular to atomic superionic water under extreme confinement and pressure}
\title{Supplementary Information: \mytitle}

\author{Ioachim G. Dusa}
\email{igd27@cam.ac.uk}
\affiliation{Cavendish Laboratory, Department of Physics, University of Cambridge, Cambridge CB3 0HE, UK}
\affiliation{Lennard-Jones Centre, University of Cambridge, Trinity Lane, Cambridge CB2 1TN, UK}

\author{Chris J. Pickard}
\affiliation{Department of Materials Science \& Metallurgy, University of Cambridge, 27 Charles Babbage Road, Cambridge CB3 0FS, UK}
\affiliation{Lennard-Jones Centre, University of Cambridge, Trinity Lane, Cambridge CB2 1TN, UK}

\author{Angelos Michaelides}
\affiliation{Yusuf Hamied Department of Chemistry, University of Cambridge, Lensfield Road, Cambridge CB2 1EW, UK}
\affiliation{Lennard-Jones Centre, University of Cambridge, Trinity Lane, Cambridge CB2 1TN, UK}

\author{Christoph Schran}
\email{cs2121@cam.ac.uk}
\affiliation{Cavendish Laboratory, Department of Physics, University of Cambridge, Cambridge CB3 0HE, UK}
\affiliation{Lennard-Jones Centre, University of Cambridge, Trinity Lane, Cambridge CB2 1TN, UK}

\date{\today}
\maketitle

\setcounter{tocdepth}{2}
\tableofcontents

\section{MACE MLP construction and error evaluation}

In this section we present the training workflow of the employed MACE MLP together with the yielded errors across the entire sampled pressure region (10 bar to 550 GPa), on both validation and training datasets. 
The trained model uses 128 scalar hidden channels, corresponding to an invariant architecture.
The model is trained with a cutoff radius of \(6.0~\text{\AA}\) and two message-passing rounds, rendering the effective cutoff radius to be \(12.0~\text{\AA}\).

As a \textbf{zeroth step} in the model development, we train a model on the dataset from \cite{Kapil2022}, containing water configurations labelled with revPBE0-D3. This dataset includes bulk ice structures, confined water configurations such as monolayers and bilayers, and interface structures, all spanning an intermediate pressure regime up to $12$ GPa.
Moreover, it builds upon that of \cite{Schran2020}, where Behler–Parrinello neural network committees were used, along with an active learning strategy to iteratively improve the neural network potential over multiple generations. 
In addition, classical and path integral NPT simulations of monolayer and bilayer water, bulk water, and hexagonal ice, as well as pressure ramp simulations of monolayer water are included.

In the \textbf{first step}, we use the MACE-OMAT foundation model~\cite{MACEMP0} to generate initial high-pressure configurations. MACE-OMAT is a pre-trained model based on data released by Meta AI~\cite{omat-dataset}, comprising DFT single-point calculations, structural relaxations and molecular dynamics trajectories for a broad range of inorganic bulk materials. The dataset contains approximately 118 million structures labelled with total energies, forces and stresses.

Using the foundation model, we sample configurations between $P = 60$ and $140~\mathrm{GPa}$ at $T = 900~\mathrm{K}$. 
These structures are relabelled using DFT and added to the training set, allowing us to take a significant pressure ``leap'' in the phase diagram. In addition, we include configurations at $30~\mathrm{GPa}$ generated by the zeroth-step model trained on the initial dataset. This pressure lies close to the regime where the zeroth model begins to lose stability, therefore, including these structures improves robustness across the low- to intermediate-pressure transition.

In the \textbf{second step}, we use the first-step model to sample the phase diagram more extensively between $P = 60$ and $140~\mathrm{GPa}$ and $T = 600$--$900~\mathrm{K}$. These configurations are again relabelled with DFT and incorporated into the training set.

Subsequent models are generated through a simple iterative learning procedure. At each iteration, the current model is used to sample at progressively higher pressures until signs of instability or reduced accuracy appear. Configurations from this limiting regime are then relabelled with DFT and added to the training set. Repeating this procedure gradually extends the pressure range over which the model is stable and accurate. In total, we perform ten iterative retraining steps, yielding a \textbf{final model} that remains reliable up to $550~\mathrm{GPa}$ and beyond. The final training data set comprises 6000 structures.

\begin{figure*}[htbp]
    \centering
    \includegraphics[width=\textwidth]{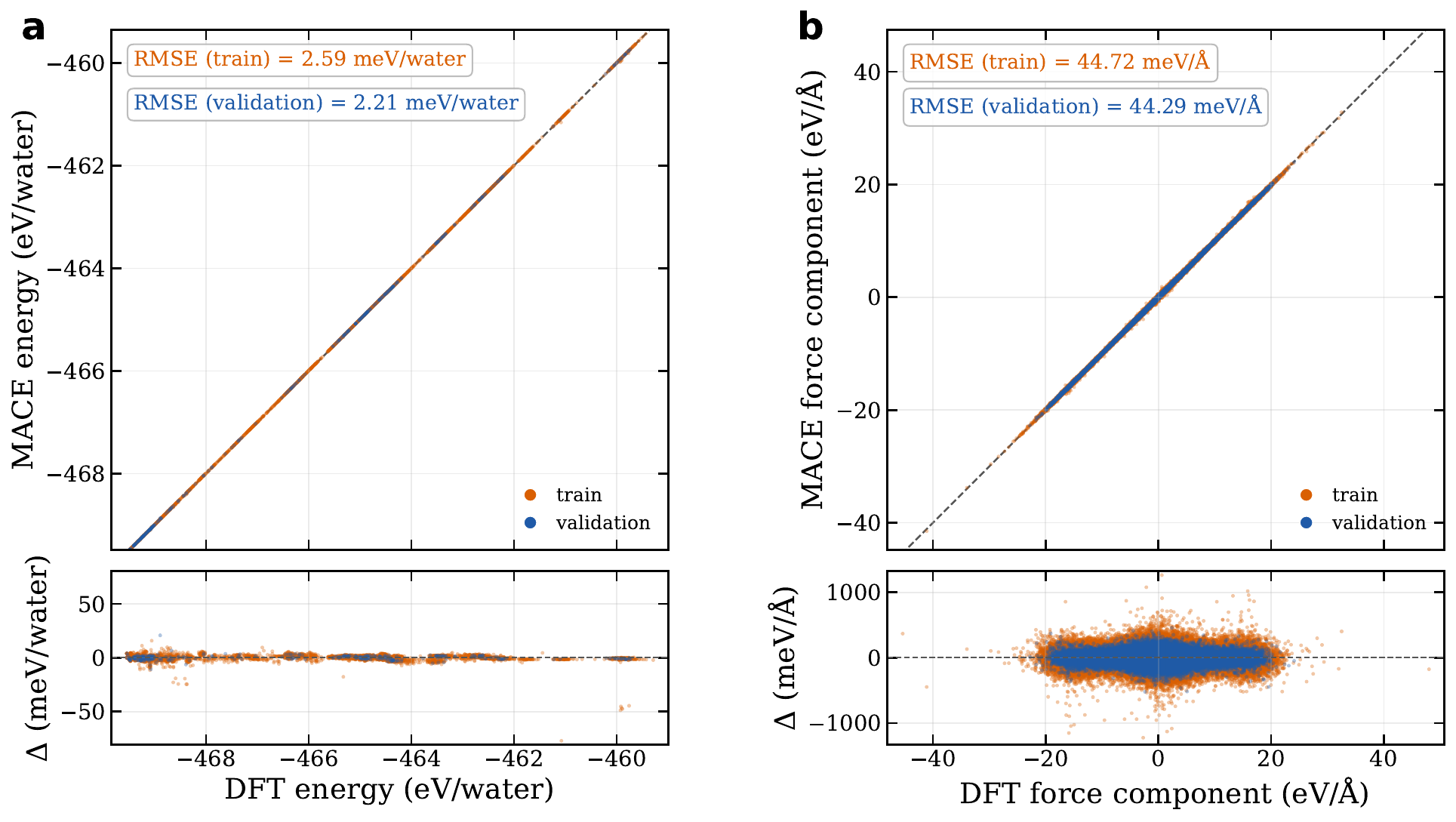}
    \caption{\textbf{Final MACE model accuracy on validation (blue) and training (orange) set: energies and forces.}
    \\
    (a) Parity plot of MLP-predicted versus DFT reference energies per
    water molecule at revPBE0-D3 level of theory. The lower panel shows
    the residual $\Delta = E_{\mathrm{MLP}} - E_{\mathrm{DFT}}$ per water
    molecule. (b) Parity plot of MLP-predicted versus DFT
    reference forces. The lower panel shows the residual
    $\Delta = F_{\mathrm{MLP}} - F_{\mathrm{DFT}}$.}
    \label{fig:S4}
\end{figure*}

The final model is assessed by evaluating the energy and force RMSEs on
the training set and validation set (Fig.~\ref{fig:S4}), as well as on configurations generated by
the model itself through random structure searches, Table~\ref{tab:spearman}, and finite-temperature molecular dynamics
simulations in Table~\ref{tab:MD-1250K}.
We further quantify the model's ability to reproduce the energetic ordering of low-enthalpy candidates via the Spearman rank correlation, $\rho$ (Table~\ref{tab:spearman}). $\rho$ exceeds 0.90 up to 400~GPa and stays above 0.85 across the full range (0.001--500~GPa), confirming very good agreement with DFT ordering throughout.

\begin{table}[htb]
\centering
\caption{Spearman ranking $\rho$ and errors of the MACE MLP predictions against DFT for the candidate ice phases at different pressures. $N$ denotes the number of irreducible ice structure candidates.}
\label{tab:spearman}
\begin{tabular}{ccccc}
\hline
$P$ (GPa) & $N$ & $E_{\mathrm{RMSE}}$ (meV/H$_2$O) & $F_{\mathrm{RMSE}}$ (meV/\AA) & $\rho$ \\
\hline
0.001 & 306 & 2.60 & 24.61 & 0.979 \\
12 & 169 & 5.53 & 26.02 & 0.975 \\
25 & 179 & 9.24  & 28.28  & 0.961\\ 
50 & 26 & 2.35 & 26.60 & 0.988\\
100 & 34 & 5.84 & 33.29 & 0.979 \\
200 & 34 & 3.72 & 36.30 & 0.958 \\
300 & 55 & 9.10 & 63.13 & 0.937 \\
400 & 16 & 5.41 & 62.52 & 0.900 \\
500 & 7 & 2.14 & 57.47 & 0.857 \\
\hline
\end{tabular}
\end{table}

\begin{table}[htb]
\centering
\caption{Energy and force RMSE of the MACE MLP with respect to DFT for statistically independent MD configurations at $1000~\mathrm{K}$.}
\label{tab:MD-1250K}
\begin{tabular}{ccc}
\hline
$P$ (GPa) & $E_{\mathrm{RMSE}}$ (meV/H$_2$O) & $F_{\mathrm{RMSE}}$ (meV/\AA) \\
\hline
100 & 2.635  & 48.078\\
200 & 4.960 & 55.226\\
300 & 1.528 & 36.582 \\
400 & 1.374 & 60.717 \\
500 & 0.271 & 121.25 \\
\hline
\end{tabular}
\label{tab:mlp_dft_md_1250K}
\end{table}

\FloatBarrier

\section{Additional implicitly confined structures}
Here we present additional high pressure metastable implicitly confined ice phases not shown in the main
text. Two of these, the distorted centred flat-square bilayer ice
(d-c-fSq-BI) and the molecular rhombic bilayer ice (m-Rh-BI), are shown
in Fig.~\ref{fig:S3}. In Fig.~2a of the main text they are close in enthalpy to each other and lie $\sim$10--25~meV/H$_2$O above c-fSq-BI between 100 and 225~GPa.

\begin{figure}[htbp]
    \centering
    \includegraphics[width=\columnwidth]{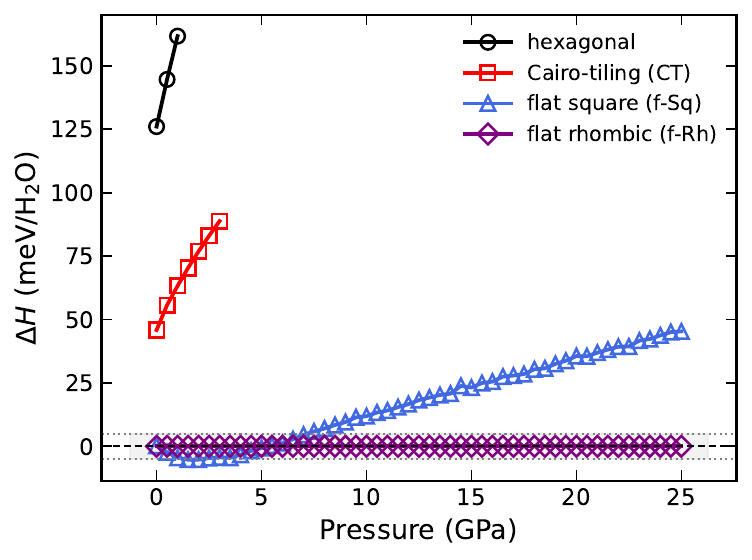}
    \caption{\textbf{Low-pressure phases of nanoconfined water ice: relative enthalpy.}
    \\
    Relative enthalpy $\Delta H$ per water molecule at $T = 0$~K
    and slit width $w = 5$~\AA{} in the low-pressure regime (0--25~GPa),
    given relative to the flat-rhombic phase (f-Rh). The shaded band
    indicates the $\pm 5$~meV/H$_2$O degeneracy window. The sampled
    candidates comprise hexagonal, Cairo-tiling (CT), flat-square (f-Sq),
    and flat-rhombic (f-Rh) monolayer ice. Corresponding structures are
    shown in Fig.~\ref{fig:S1b}.}
    \label{fig:S1a}
\end{figure}

\begin{figure}[htbp]
    \centering
    \includegraphics[width=0.9\columnwidth]{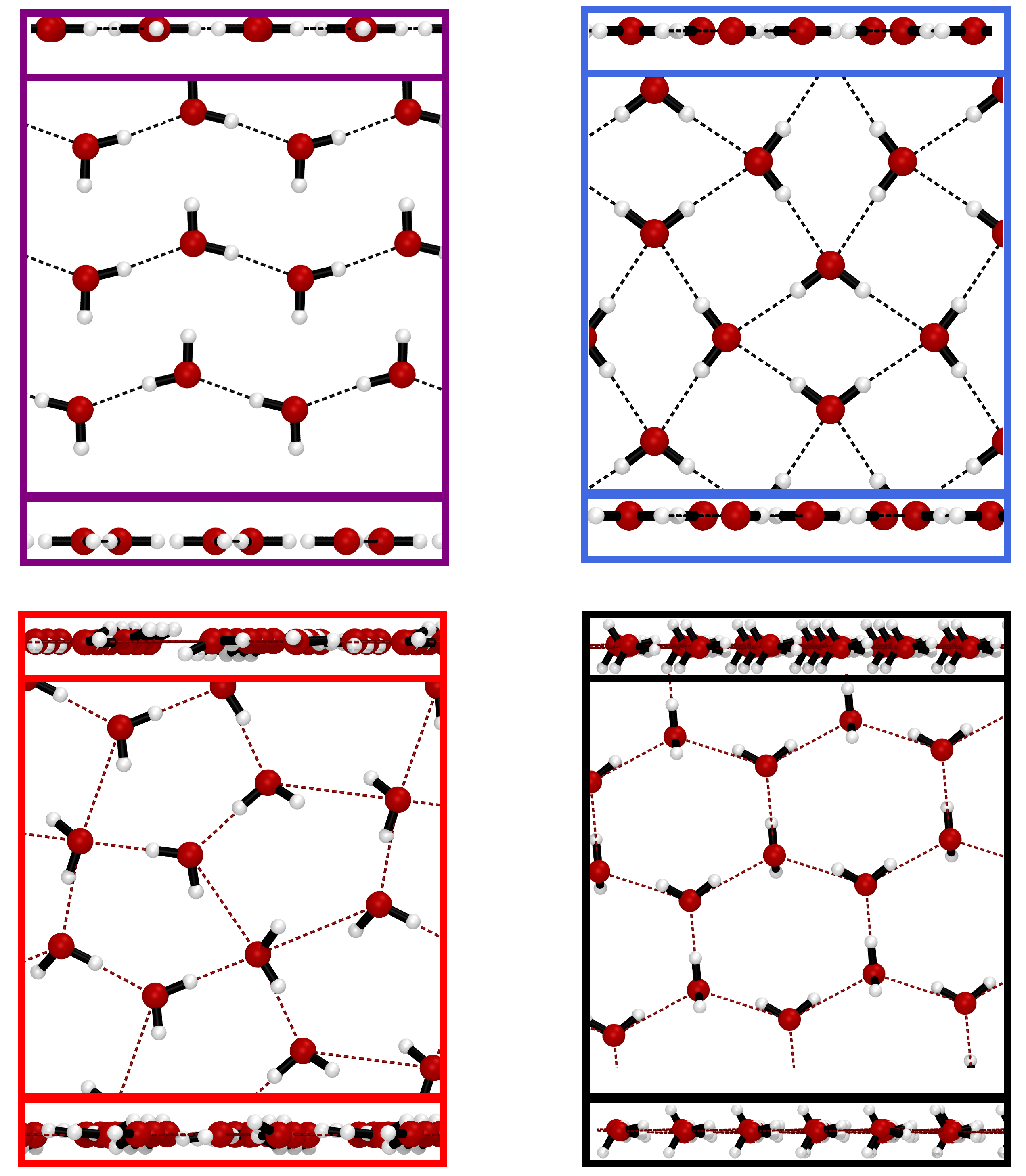}
    \caption{\textbf{Low-pressure phases of nanoconfined water ice: structures.}
    \\
    Structures corresponding to the enthalpy curves in Fig.~\ref{fig:S1a},
    shown in top view (centre) and side view (top and bottom panels of
    each frame): flat-rhombic (purple), flat-square (blue), Cairo-tiling
    (red), and hexagonal (black). Red spheres denote oxygen, white
    spheres hydrogen; dashed lines indicate hydrogen bonds.}
    \label{fig:S1b}
\end{figure}

Additionally, we report the ice phases sampled in the low-pressure
regime in Fig.~\ref{fig:S1a} and Fig.~\ref{fig:S1b}, where the
flat-square and flat-rhombic phases are found to be the most stable,
as opposed to the previously predicted hexagonal and Cairo-tiling
phases, for example in Ref.~\cite{2DIce-ab-init}. This is mainly
attributed to the hydrogen-confining potential introduced in our
setup, which presses the hydrogens into the confinement plane, making
the outward-pointing hydrogens of the Cairo-tiling and hexagonal
phases energetically unfavourable.

\begin{figure}[htbp]
    \centering
    \includegraphics[width=\columnwidth]{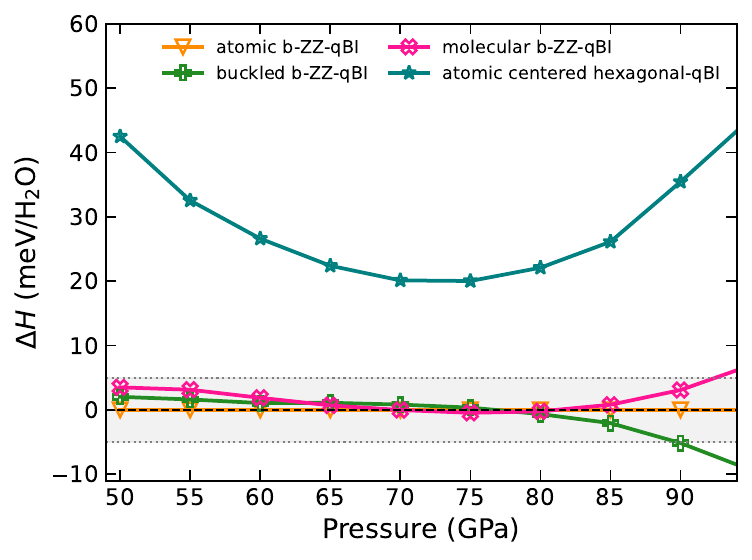}
    \caption{\textbf{Intermediate-pressure phases of nanoconfined water ice: relative enthalpy.} 
    \\
    Relative enthalpy $\Delta H$ per water molecule at
    $T = 0$~K and $w = 5$~\AA{} in the intermediate-pressure regime
    (25--100~GPa), given relative to the atomic branched zig-zag quasi-bilayer
    phase (b-ZZ-qBI). The shaded band indicates the $\pm 5$~meV/H$_2$O
    degeneracy window. Corresponding structures are shown in
    Fig.~\ref{fig:S2b}.}
    \label{fig:S2a}
\end{figure}

\begin{figure}[htbp]
    \centering
    \includegraphics[width=0.89\columnwidth]{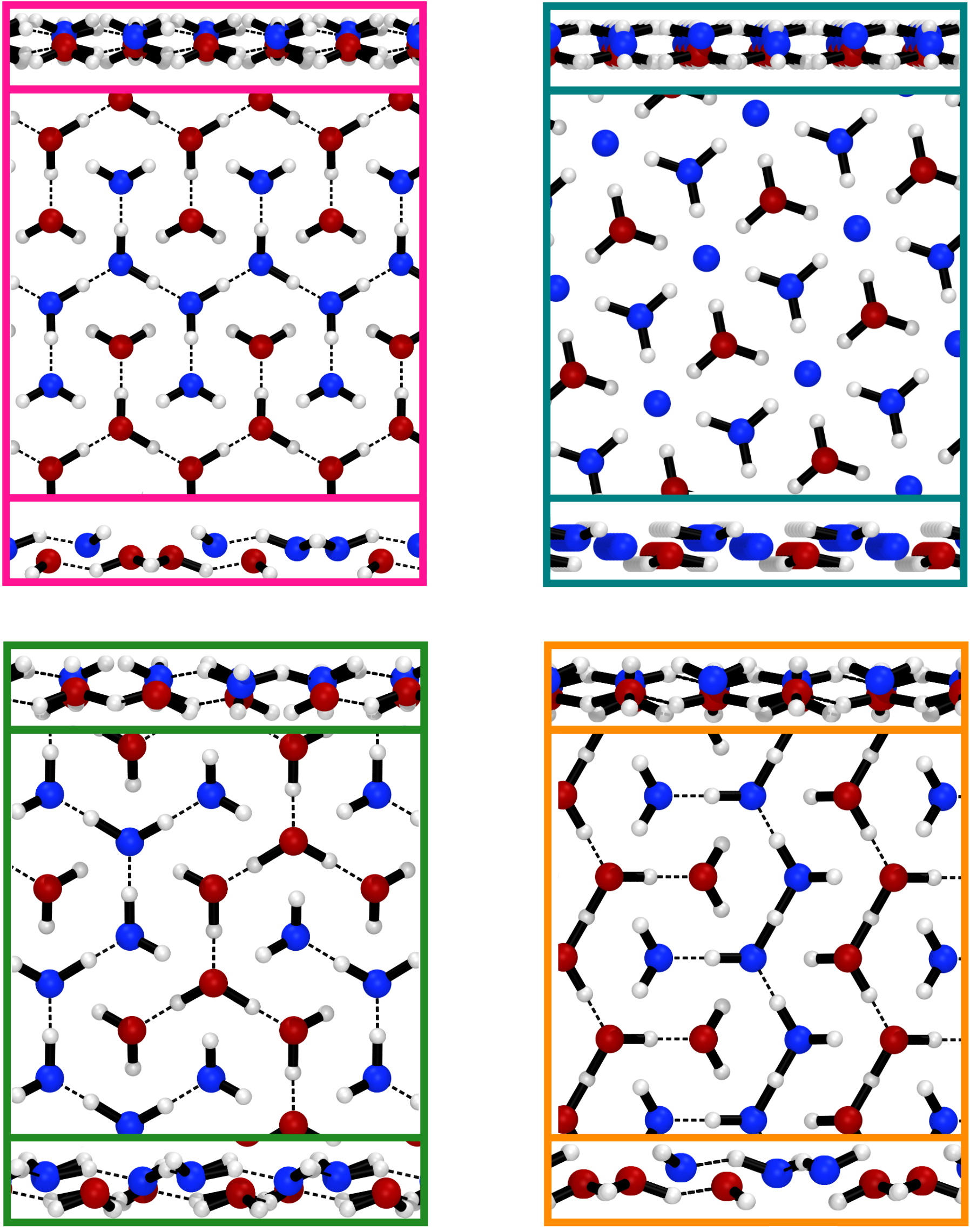}
    \caption{\textbf{Intermediate-pressure phases of nanoconfined water ice: structures.} 
    \\
    Structures corresponding to the enthalpy curves in Fig.~\ref{fig:S2a},
    shown in top and side views, with frame colours matching the legend}
    \label{fig:S2b}
\end{figure}

\begin{figure}[htbp]
    \centering
    \includegraphics[width=0.845\columnwidth]{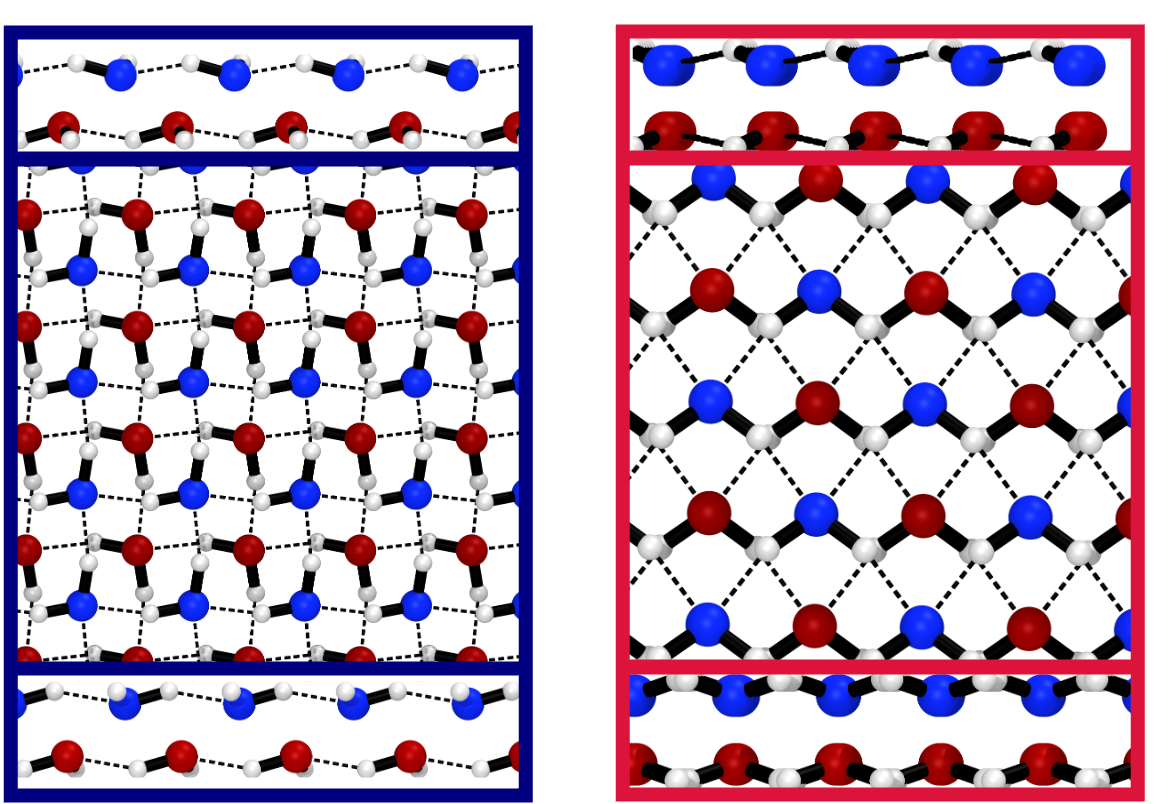} 
    \caption{\textbf{Metastable high-pressure bilayer structures.} 
    \\
    Side (top and bottom panels) and top views (centre panel) of the two
    metastable bilayer candidates appearing in the high-pressure enthalpy
    ordering of Fig.~2a of the main text: the distorted centred flat-square
    bilayer ice (d-c-fSq-BI, blue frame) and the molecular rhombic bilayer
    ice (m-Rh-BI, red frame). Blue and red spheres denote oxygen atoms in
    the two different sheets stacked along the confinement direction
    $\hat{z}$; white spheres denote hydrogen; dashed lines indicate
    H--O bonding. The corresponding enthalpy curves are shown in the
    main text (Fig.~2a).}
    \label{fig:S3}
\end{figure}

Moreover, we also sample the intermediate-pressure regime, shown in
Fig.~\ref{fig:S2a} and Fig.~\ref{fig:S2b}, where the b-ZZ-qBI
phase~\cite{jiang_rich_2024} is found to be the dominant structure,
alongside several closely related structural variants.

The structural data for every phase presented here and in the main
text are provided as extended XYZ files in the \texttt{ice-structures/}
directory of the associated GitHub repository:
\href{https://github.com/fast-group-cam/high-p-phase-transitions}{https://github.com/fast-group-cam/high-p-phase-transitions}.

\FloatBarrier

\section{ZPE calculations}
\begin{figure*}[htbp]
    \centering
    \includegraphics[width=0.95\textwidth]{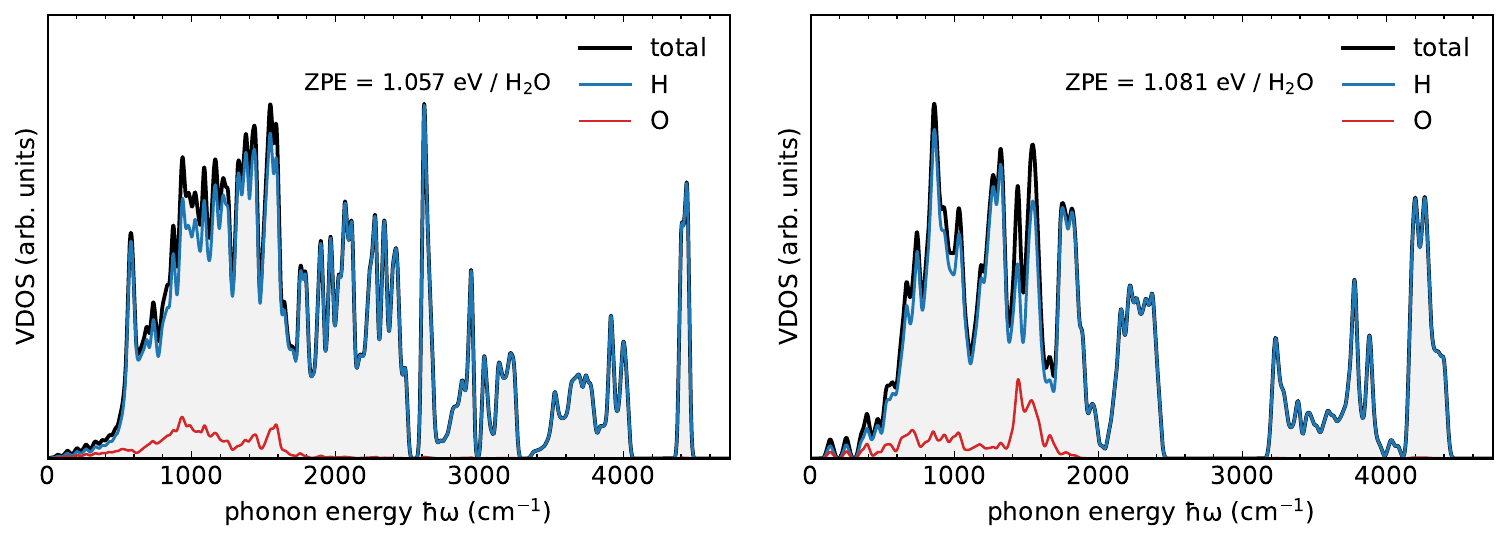}
    \caption{\textbf{vDOS and zero-point energy of the high pressure candidates at $T=0$~K and $P=500$~GPa, computed using the MACE model.} \\
    Left panel: a-Ch-BI phase. Right panel: a-Rh-BI phase.}
    \label{fig:S6}
\end{figure*}

To assess the magnitude of the zero-point energy corrections of high pressure structures, we perform harmonic phonon computations for a-Rh-BI and a-Ch-BI using the MACE model employed in this work and the finite-displacement method implemented in the python package \texttt{phonopy}~\cite{phonopy-phono3py-JPCM, phonopy-phono3py-JPSJ}.
For both structures, the vibrational density of states (VDOS) is dominated by the hydrogens.
Moreover, the main qualitative difference between a-Ch-BI (Fig.~\ref{fig:S6}, left panel) and a-Rh-BI (Fig.~\ref{fig:S6}, right panel) is the appearance of additional spectral weight in the intermediate-frequency region around \(2500\)--\(3200~\mathrm{cm^{-1}}\) for a-Ch-BI.
This feature is explained by softer O--H stretching modes arising from the more connected, interpenetrating bonding network, whereas the corresponding frequency window remains largely empty in a-Rh-BI.
The species-resolved contributions to the VDOS are obtained by projecting each phonon eigenvector onto hydrogen and oxygen degrees of freedom. In both phases, the hydrogen-projected contribution accounts for the dominant part of the total ZPE, with \(E_\mathrm{ZPE}\) of order \(\sim1~\mathrm{eV/water}\).

\FloatBarrier

\section{Free energy profiles}

In this section, we present free energy profiles resolved with
respect to the oxygen--oxygen separation $r_\text{OO}$, the proton
transfer coordinate $\delta$, and the O--H bond orientation, the
latter parametrized by the azimuthal angle $\phi$ within the
confining ($xy$) plane in the implicitly confined case. The proton transfer coordinate $\delta$ is
defined such that $\delta = 0$ corresponds to a proton lying
perfectly equidistant between two oxygen atoms.

\begin{figure*}[htbp]
    \centering
    \includegraphics[width=\textwidth]{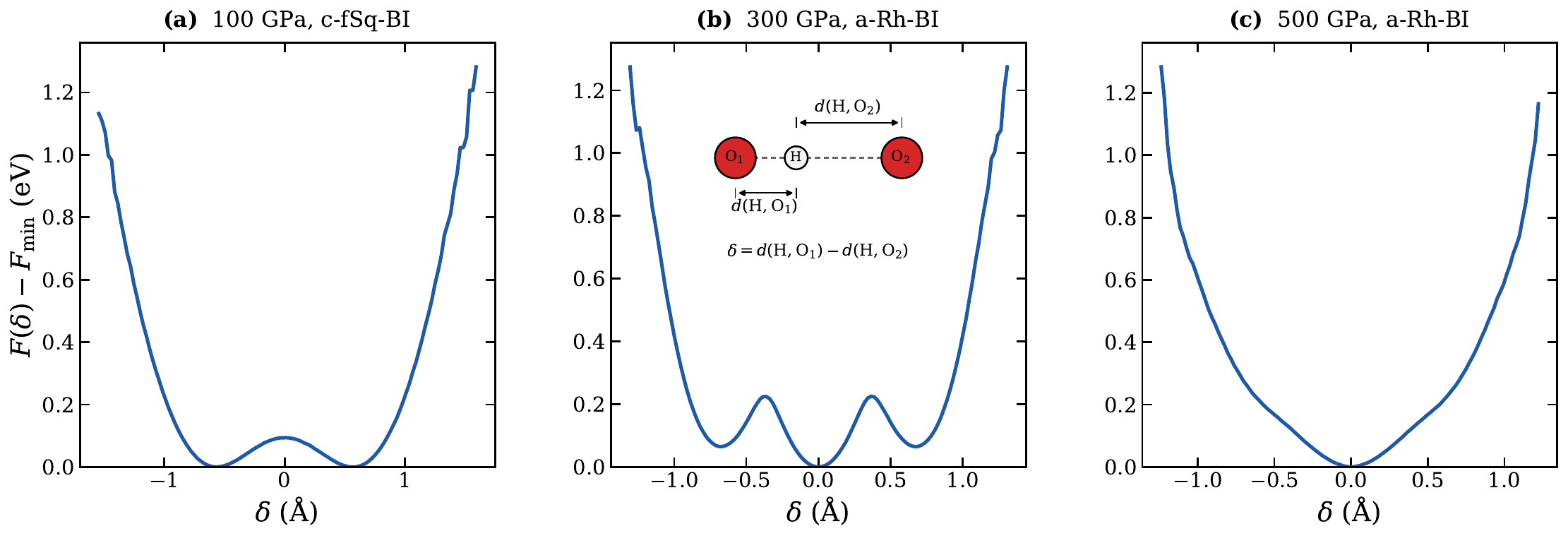}
    \caption{\textbf{Free energy along the proton-transfer coordinate $\delta$ at $T = 1250$~K.} \\
    One-dimensional profiles $F(\delta)$ for the three implicitly confined phases, ordered by increasing
    pressure from left to right: (a) $100$~GPa, c-fSq-BI; (b) $300$~GPa, a-Rh-BI; (c) $500$~GPa, a-Rh-BI.
    The sketch in (b) illustrates the definition of the coordinate: for each H atom, the two nearest
    oxygens O$_1$ and O$_2$ are identified and $\delta = d(\mathrm{H,O_1}) - d(\mathrm{H,O_2})$ is the
    difference of the two H--O distances along O--H$\cdots$O. Thus $\delta = 0$ corresponds to a proton
    equidistant between the two oxygens, while $\delta < 0$ ($\delta > 0$) indicates a proton bonded
    to O$_1$ (O$_2$).}
    \label{fig:fe_delta}
\end{figure*}

\begin{figure*}[htbp]
    \centering
    \includegraphics[width=\textwidth]{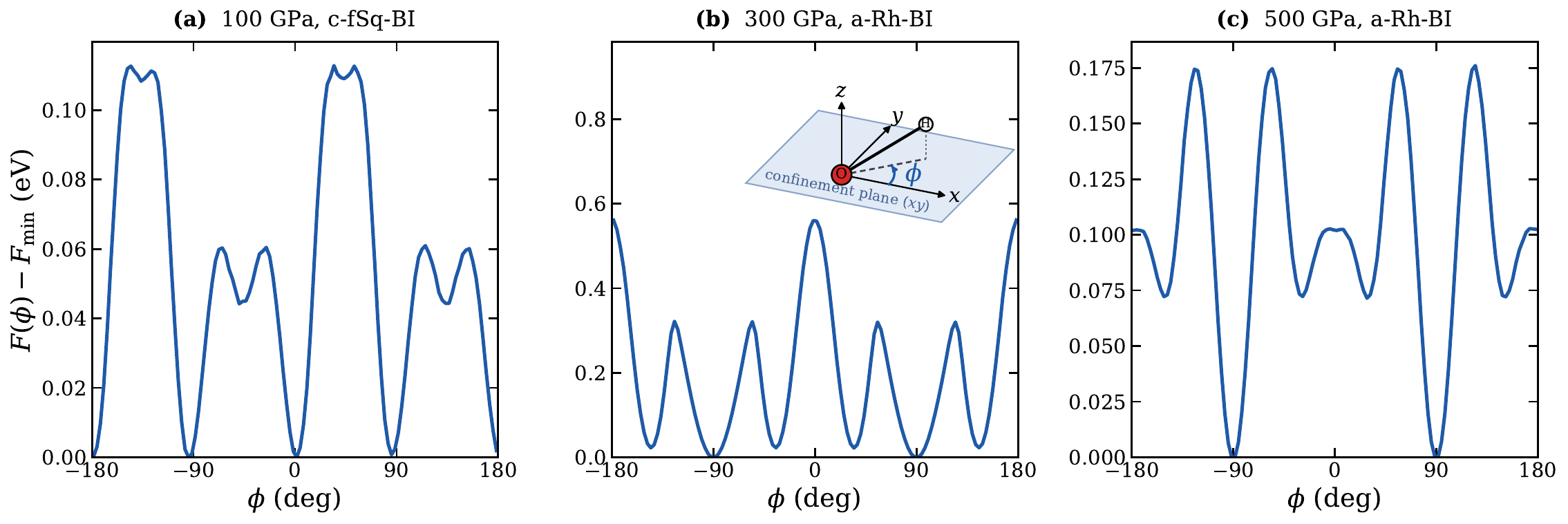}
    \caption{\textbf{Free energy along the azimuthal angle $\phi$ of the O--H bond at $T = 1250$~K.} \\
    One-dimensional profiles $F(\phi)$ for the three implicitly confined phases, ordered by increasing
    pressure from left to right: (a) $100$~GPa, c-fSq-BI; (b) $300$~GPa, a-Rh-BI; (c) $500$~GPa, a-Rh-BI.
    The sketch in (b) illustrates the definition of $\phi$: the O--H bond vector $\mathbf{v}$ (solid
    line) may tilt out of the confinement plane ($xy$, shaded); its projection onto that plane
    (dashed line) defines the azimuthal angle $\phi = \mathrm{atan2}(v_y, v_x)$, measured from the
    $x$-axis. $\phi$ therefore describes the in-plane orientation of the O--H bond.}
    \label{fig:fe_phi}
\end{figure*}

We compute these quantities at $1250$~K for the three phases characteristic of the conductivity transitions identified in Fig.~2b of the main text,
namely the c-fSq-BI phase at $100$~GPa and the a-Rh-BI phase at
$300$~GPa and $500$~GPa (Fig.~\ref{fig:fe_delta} and Fig.~\ref{fig:fe_phi}). The a-Rh-BI phase exhibits higher
free-energy barriers at 300 GPa than c-fSq-BI at 100 GPa, most notably for O--H bond
rotation (along $\phi$) but also for proton hopping (along
$\delta$), consistent with the reduced conductivity of a-Rh-BI
at the first conductivity transition. Notably, at 300 GPa the $\delta$ profile of the a-Rh-BI
phase displays three distinct minima rather than the conventional
double-well form: in addition to the two outer minima corresponding
to a proton localized on either oxygen, a third, deeper minimum
appears at $\delta = 0$. This indicates that the proton spends a
non-negligible amount of time resting in a symmetric configuration
midway between the two oxygens, before being pulled towards one of them.

\begin{figure*}[htbp]
    \centering
    \includegraphics[width=0.60\textwidth]{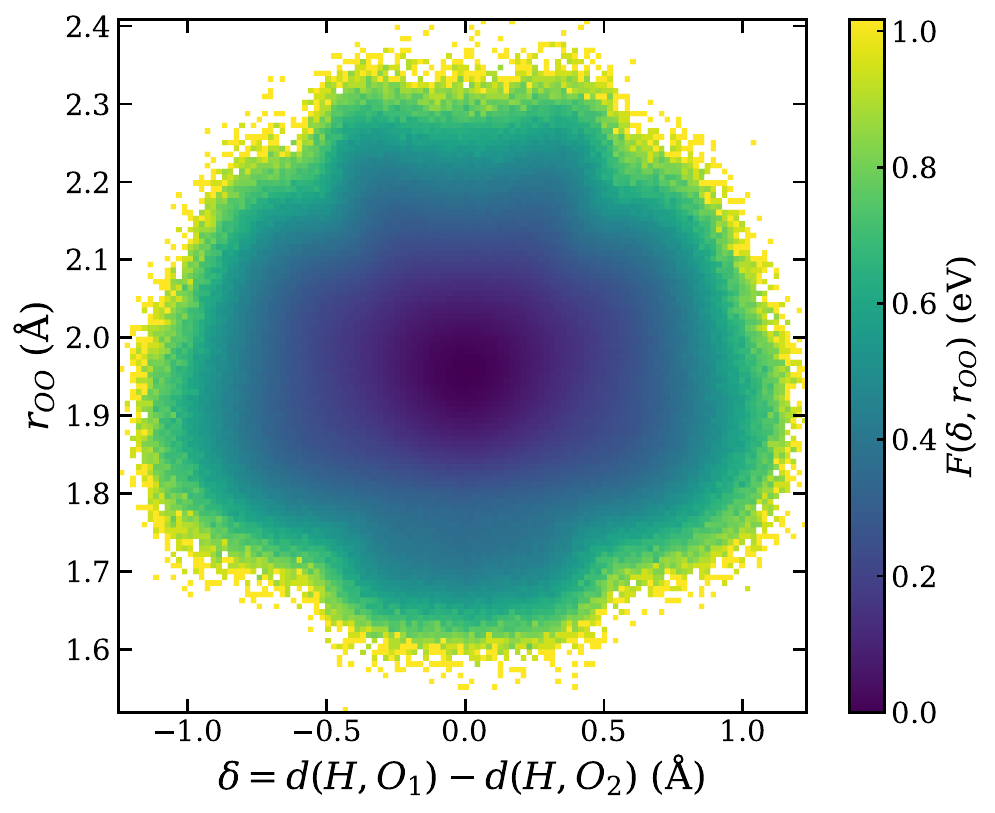}
    \caption{\textbf{Two-dimensional free energy surface $F(\delta, r_\mathrm{OO})$ for the a-Rh-BI
    phase at $T = 1250$~K and $P = 500$~GPa.} \\
    Free energy as a function of the proton-transfer coordinate $\delta$ and the oxygen--oxygen
    separation $r_\text{OO}$ under implicit confinement. As in Fig.~\ref{fig:fe_delta}, $\delta = 0$ corresponds to a proton
    equidistant between the two oxygens separated by $r_\text{OO}$.}
    \label{fig:fe_2d_500}
\end{figure*}

In addition, we compute a two dimensional free energy profile along $\delta$ and $r_\text{OO}$, for the a-Rh-BI phase at $500$~GPa and $1250$~K
(Fig.~\ref{fig:fe_2d_500}). At this higher pressure, the profile
displays a single well, indicating that the transferring proton sits, on
average, exactly midway between two oxygen atoms, which suggests that
proton transport throughout the lattice proceeds predominantly via
O--H bond rotation rather than hopping between oxygens --
in contrast to the c-fSq-BI phase, where transport is governed
jointly by both rotation and proton transfer between adjacent oxygen
atoms.

\section{Additional RDFs}

\begin{figure*}[htbp]
    \centering
\includegraphics[width=\textwidth]{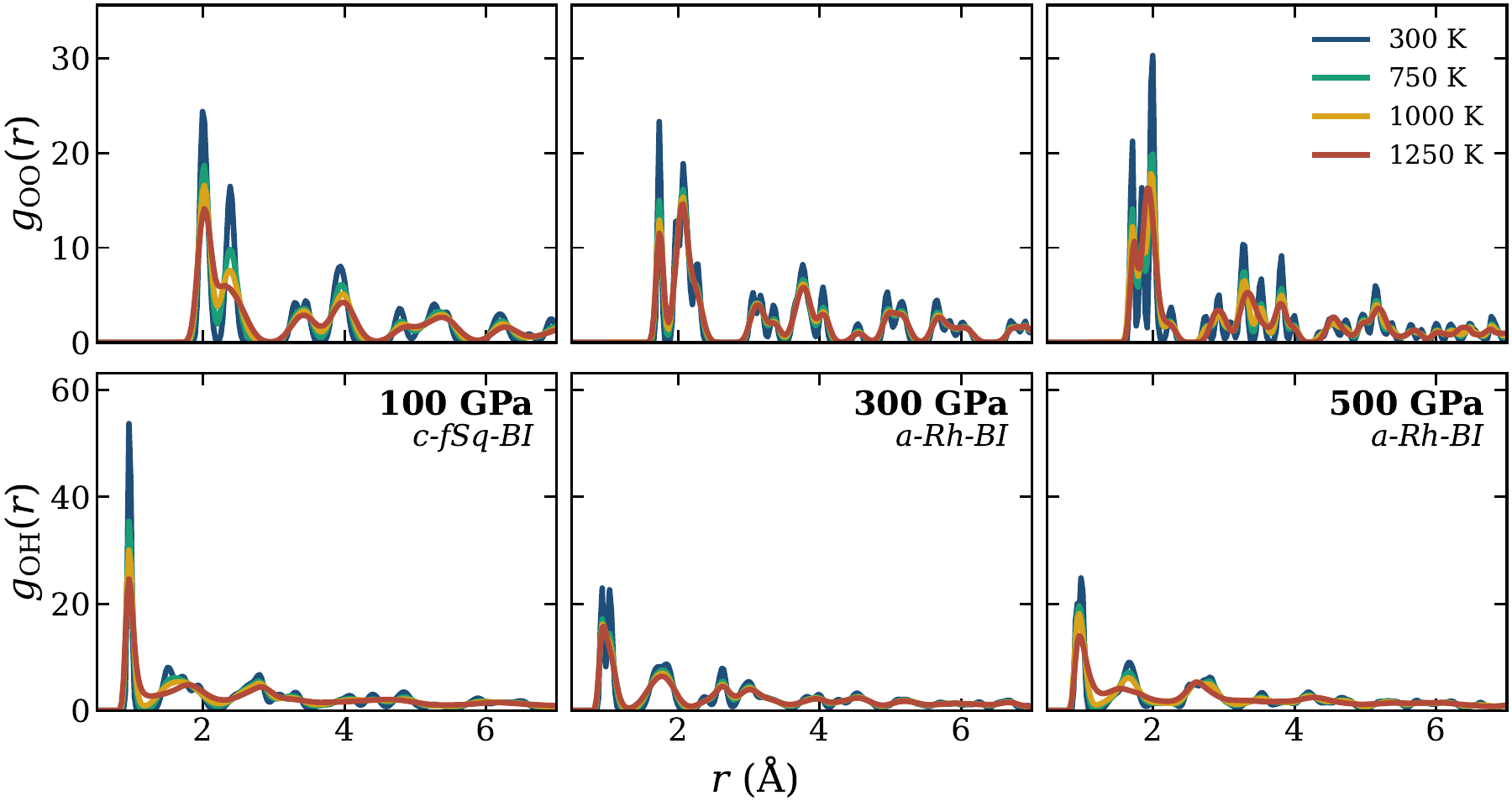}
    \caption{\textbf{Oxygen--oxygen and oxygen--hydrogen radial
    distribution functions across temperature and pressure under implicit confinement.} \\ $g_{\mathrm{OO}}(r)$
    (top panels) and $g_{\mathrm{OH}}(r)$ (bottom panels) at $300$,
    $750$, $1000$, and $1250$~K for the c-fSq-BI phase at
    $100$~GPa, the a-Rh-BI phase at $300$~GPa, and the a-Rh-BI
    phase at $500$~GPa.}
    \label{fig:S10}
\end{figure*}

We present additional radial distribution functions at $100$, $300$,
and $500$~GPa under implicit confinement, showing the O--O and O--H pair distribution functions
across a range of temperatures ($300$--$1250$~K), providing further
insight into the lattice structure throughout the sampled
pressure range (Fig.~\ref{fig:S10}).

At $100$~GPa, in the superionic regime of the c-fSq-BI phase
(Fig.~\ref{fig:S10}a), the hydrogen $g_{\mathrm{OH}}(r)$ structure
departs from that of a solid hydrogen lattice and instead resembles a
liquid-like one, evidenced by increasing peak broadening and the
merging of adjacent peaks as temperature increases. We observe the
same behaviour at $500$~GPa in the superionic regime of the a-Rh-BI
phase (Fig.~\ref{fig:S10}c), where the corresponding peaks likewise
broaden and overlap at high temperature. This is in contrast to the
$300$~GPa case (non-superionic), again the a-Rh-BI phase (Fig.~\ref{fig:S10}b), where
the solid-like hydrogen lattice structure is preserved even at
$1250$~K, with no peak overlap occurring.

\FloatBarrier

\section{Explicit diamond anvil cell setup}
\begin{figure}[htbp]
    \centering
\includegraphics[width=0.7\columnwidth, height=0.71\textheight, keepaspectratio]{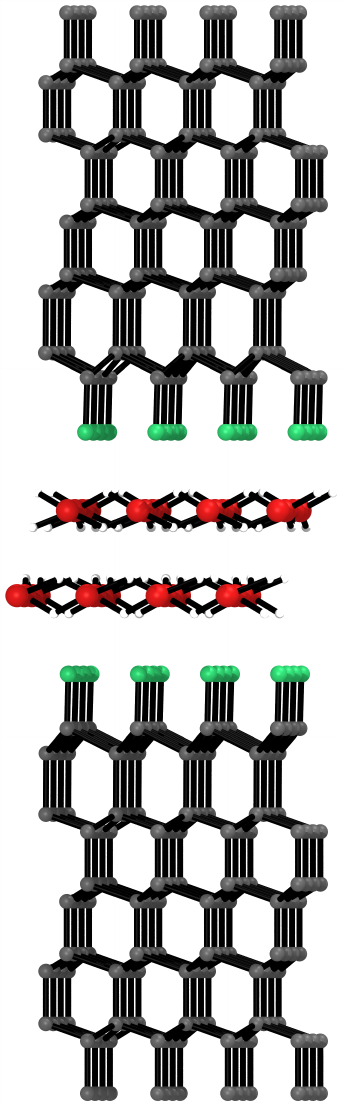}
    \caption{\textbf{Diamond-anvil-cell confinement geometry.} \\
    Relaxed water structure confined between fluorine-terminated diamond
    surfaces. The confining walls are constructed by cleaving diamond along
    the (111)-surface and fully terminating the exposed surface atoms with fluorine.
    The full water--wall structure is subsequently relaxed. Red and white
    spheres denote oxygen and hydrogen atoms, respectively; grey and green
    spheres denote carbon and fluorine atoms in the confining diamond slabs. In this example the fluorine atoms on opposing sides of the confining slabs are separated by $5.7$~\AA{}. The simulation cell used in the ab~initio computations contains 120 atoms, including eight water
    molecules, and is shown as a $2\times2$ in-plane supercell for
    visualisation purposes.}
    \label{fig:diamond-anvil}
\end{figure}

To scrutinize the effect of explicit confinement on nanoconfined
water, we construct a realistic setup potentially resembling
diamond-anvil-cell experiments (Fig.~\ref{fig:diamond-anvil}). We
choose the confining surface to be the (111)-surface of diamond,
fully terminated with fluorine, as this has been shown in the
literature to be a chemically stable termination~\cite{FterminatedD}.

The workflow is carried out in three stages. We first relax bulk
diamond at $\sim 600$~GPa to extract its lattice
parameter. We then
construct the diamond anvil by cleaving bulk diamond along the (111)-surface, terminating the exposed surface with fluorine, and
performing a geometry optimization of the bare diamond slab (without
water). 
Water is then introduced at the areal density of the implicitly confined bilayer at $\sim$500~GPa, allowing a maximum strain of 3\% to fit within the diamond geometry and the resulting structure is relaxed.
In the final relaxation, only the water, the fluorine termination, and the first two atomic layers of diamond on each side of the confinement direction are allowed to relax, keeping the computational cost tractable. We sample
separations between the fluorine terminations on opposing sides of
the confining slabs, $d_{\mathrm{F-F}}$, ranging from $5.7$ to
$6.4$~\AA{}. All geometries contain 120 atoms (including eight water molecules) in a hexagonal cell with in-plane lattice parameters $a = b = 4.03$~\AA{}
($\gamma = 60^\circ$) and an out-of-plane dimension of the overall cell $c \sim 27$~\AA{}, varying slightly with the chosen slit width.
Geometry optimizations and electronic structure calculations are performed at the PBE level of theory. All stages employ the FHI-aims code. We use the intermediate basis set for the bulk diamond lattice extraction, the light basis set for the two relaxation steps to keep computation times feasible, and the intermediate basis set again for the subsequent electronic structure calculations.

We find the sampled explicit confinement geometries to be stable not only statically, but also
under \textit{ab initio} molecular dynamics at $1250$~K in the \textit{NVT}
ensemble (see SI Video 5). Here, we sample a $10~\text{ps}$ trajectory and we observe proton hopping indicating a preserved superionic character. As in the relaxations, the computational cost is kept manageable by allowing only the water and the first few layers of the diamond anvil framework to move, while the remaining atoms are held fixed.
The MD computations are performed using the CP2K code employing the SZV basis set. The diamond-water ice structures sampled above are used as initial configurations for the MD runs.

\begin{figure*}[t]
    \centering
    \makebox[\textwidth][l]{\textbf{a}}\\
    \includegraphics[width=\textwidth]{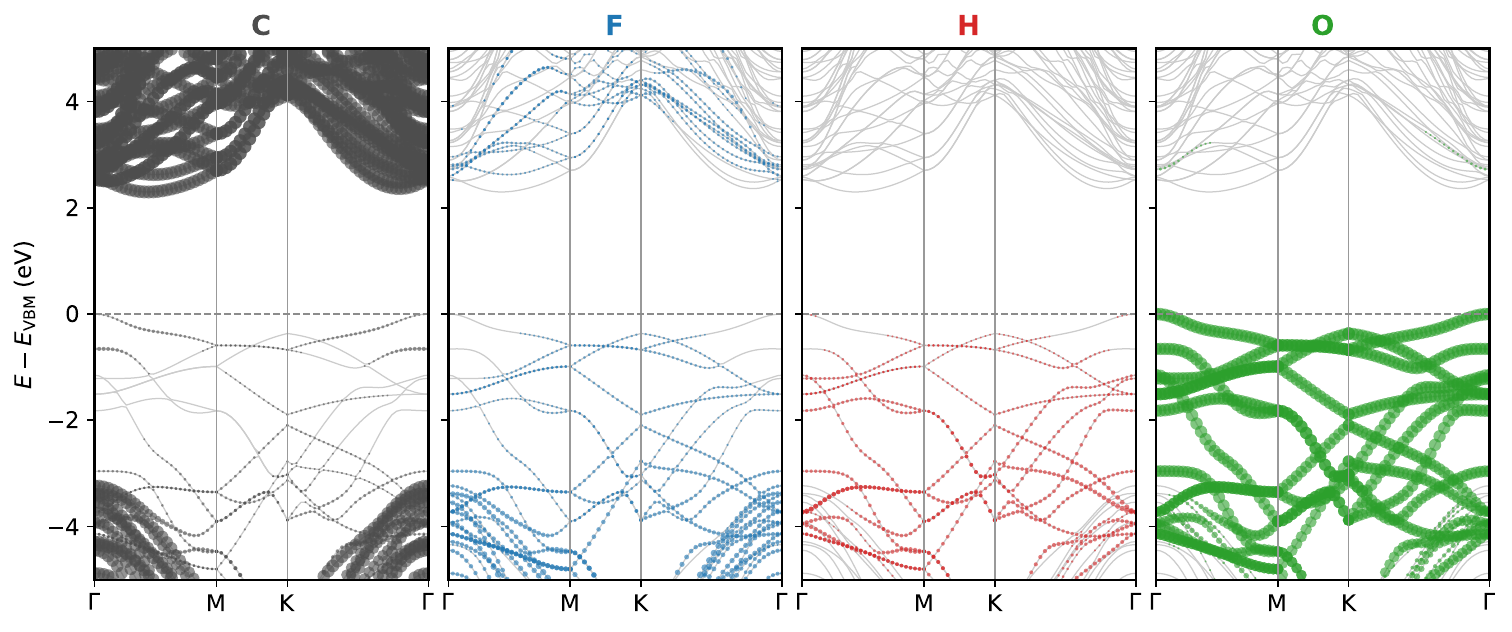}\\[2mm]
    \makebox[\textwidth][l]{\textbf{b}}\\
    \includegraphics[width=0.8\textwidth]{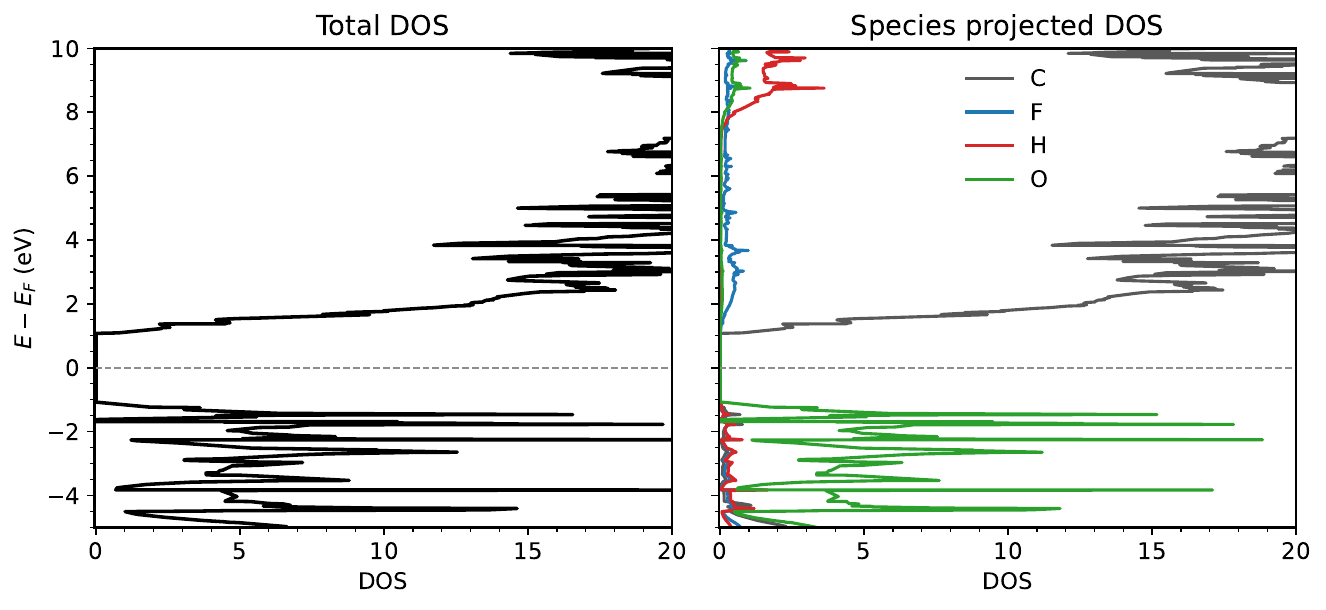}
    \caption{\textbf{Electronic band structure and density of states of the diamond-confined ice at $500$~GPa.}
    \\
    \textbf{a)} Species-projected band structure along $\Gamma$--M--K--$\Gamma$ for C, F, H and O,
    computed at the PBE level of theory for the diamond-anvil-cell structure of
    Fig.~\ref{fig:diamond-anvil}, filled with a bilayer of ice (T = 0 K) and relaxed at $500$~GPa.
    Marker size indicates the projection weight onto each species, and grey lines show
    the full band structure. Energies are given relative to the valence band maximum
    $E_\mathrm{VBM}$.
    \textbf{b)} Total (left) and species-projected (right) density of states (DOS) of the
    same structure. Energies are given relative to the Fermi level $E_F$ (dashed line).
    As in the implicit system, the valence band edge is dominated by O states (green),
    whereas the conduction band edge is now dominated by C states (grey), in stark
    contrast to the implicit case. The electronic structure computations yield a band
    gap of $\sim 2.15$~eV.}
    \label{fig:dos-diamond-anvil}
\end{figure*}

To assess the electronic structure correctly, we now let the cell relax fully, i.e.\ every atom is free to move, under an isotropic pressure of $500$~GPa again at the PBE level of theory.
Interestingly, as in the implicit case, we find that the structures exhibit a pure oxygen lone-pair character at the top of the valence band. However, the H-derived conduction states are now pushed up substantially, such that the conduction band minimum consists of pure C levels and the gap opens up by $\sim2.1$~eV at PBE level of theory (Fig.~\ref{fig:dos-diamond-anvil}).

The change in the band character, relative to the implicit-confinement case presented in the main text, arises from the band alignment of the two materials, together with the confinement of the H-derived states, which delocalise into the vacuum beyond the implicit walls but are now confined by the diamond.
The O-derived valence states of the water lie $\sim$3~eV above the C-derived valence band maximum, inside the band gap of the diamond, so that the C-derived conduction states lie directly above them, while the H-derived states are pushed $\sim$8~eV above the O-derived valence states (Fig.~\ref{fig:dos-diamond-anvil}).
Consistently, comparing the dry fluorine-terminated diamond with the full system containing the water bilayer, both relaxed at 500~GPa, the water raises the chemical potential from $\mu_{\mathrm{diamond-F-dry}} = -7.0$~eV to $\mu_{\mathrm{diamond-F-water}} = -5.8$~eV, as the highest occupied states move from the C-derived to the O-derived valence band.

To further assess the stability of the explicit confinement setup, we employ larger simulation cells using ephemeral data-derived potentials (EDDP).
The largest simulated cell contains 4000 atoms, built as a supercell of the diamond anvil geometry described above, with box dimensions of $2.1 \times 2.2 \times 2.5$~nm$^3$.
In contrast to the calculations above, every atom is allowed to move.
We simulate $0.5$~ns of trajectory in the \textit{NPT} ensemble at $1000$~K and an isotropic pressure of $650$~GPa and present a $50$~ps portion in SI Video~6.
We observe stable trajectories in which the confined water retains a crystalline oxygen lattice and its superionic character. 

The scripts, input structures, and optimized geometry can be
found at
\href{https://github.com/fast-group-cam/high-p-phase-transitions}{https://github.com/fast-group-cam/high-p-phase-transitions}.

\FloatBarrier

\section{O--O distances and volume per formula unit}

Here we examine the volume per H$_2$O formula unit and the nearest-neighbour oxygen--oxygen distance $d_\mathrm{OO}$ as a function of pressure, with $d_\mathrm{OO}$ resolved into in-plane and out-of-plane components.

\begin{figure}[htbp]
    \centering
    \includegraphics[width=\columnwidth]{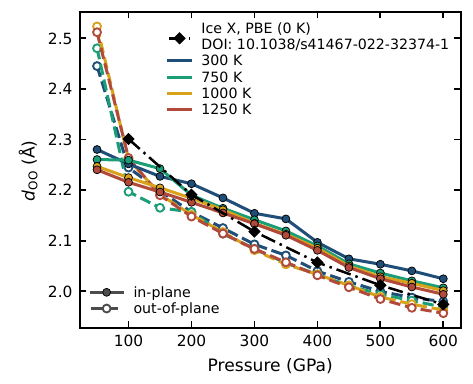}
    \caption{\textbf{Nearest-neighbour O--O distance under explicit diamond-anvil confinement.} \\
    In-plane and out-of-plane nearest-neighbour $d_\mathrm{OO}$ as a function of pressure,
    for all sampled temperatures. The black dash-dotted line
    is the $0$~K, nearest-neighbour distance of bulk ice X from
    Ref.~\cite{reinhardt_thermodynamics_2022_doo_ice_x} computed at PBE level.}
    \label{fig:dOO-explicit}
\end{figure}

For the explicit setup, we additionally compare $d_\mathrm{OO}$ to PBE bulk Ice X distances reported in Ref.~\cite{reinhardt_thermodynamics_2022_doo_ice_x}, see Fig.~\ref{fig:dOO-explicit}. Ref.~\cite{reinhardt_thermodynamics_2022_doo_ice_x} does not report $d_\mathrm{OO}$ directly, only the cubic lattice parameter $a$ of Ice X. Since the oxygen sublattice in this phase is body-centred cubic, we obtain the corresponding nearest-neighbour O--O distance via $d_\mathrm{OO} = \sqrt{3}/2 \cdot a$.
Moreover, again for the explicit setup, we report the volume per H$_2$O formula unit and compare it to the BLYP Ice X $P$--$V$ equation of state of Ref.~\cite{osmond_ultrahigh_2026_volume_blyp}, see Fig.~\ref{fig:Vwater-explicit}.
The finite-temperature explicit-confinement data track this curve closely over the reported pressure range, which gives us confidence that the explicit setup is sampling pressures comparable to bulk studies.

For the volume per formula unit under explicit confinement (Fig.~\ref{fig:Vwater-explicit}), we construct the shaded band as follows.
The height available to the confined water is not sharply defined, since it depends on where one takes the water to end and the diamond anvil to begin.
We bound it by two extremes: $h_\mathrm{F}$, measured between the fluorine termination planes, and $h_\mathrm{H}$, measured between the sheets of outward-pointing hydrogens.
Any height $h \in [h_\mathrm{H}, h_\mathrm{F}]$ gives a volume per formula unit as $h \times A/N$, where $A$ is the lateral cell area and $N$ the number of water molecules in the slit.
The solid curve uses the midpoint height, $(h_\mathrm{F} + h_\mathrm{H})/2$.
The band spans $h_\mathrm{H} + (h_\mathrm{F} - h_\mathrm{H})/4$ to $h_\mathrm{F} - (h_\mathrm{F} - h_\mathrm{H})/4$.
We exclude the outer quarter on each side because the bare F-sheet and bare H-sheet placements are themselves unlikely choices for where the water ends.

\begin{figure}[htbp]
    \centering
    \includegraphics[width=\columnwidth]{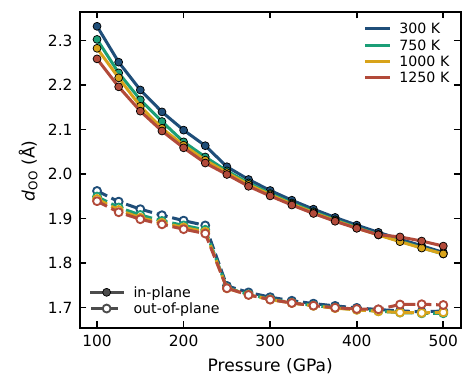}
    \caption{\textbf{Nearest-neighbour O--O distance under implicit confinement.} \\
    In-plane and out-of-plane nearest-neighbour $d_\mathrm{OO}$ as a function of pressure,
    for all sampled temperatures. The out-of-plane distance is
    systematically shorter than the in-plane one across the whole pressure
    range, reflecting the anisotropic structural response of the fixed-slit-width
    confinement discussed in the text.}
    \label{fig:dOO-implicit}
\end{figure}

\begin{figure}[htbp]
    \centering
    \includegraphics[width=\columnwidth]{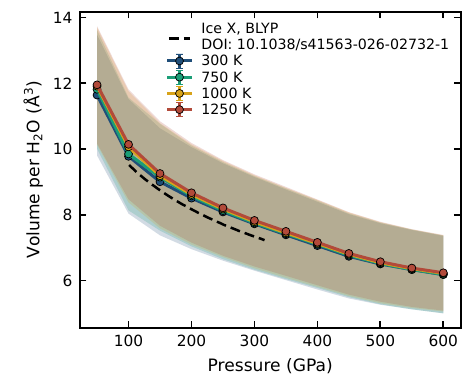}
    \caption{\textbf{Volume per H$_2$O formula unit under explicit diamond-anvil confinement.} \\
    Solid lines show the volume per H$_2$O at the midpoint between the F-sheet
    and H-sheet extremes of the assumed water/diamond boundary. Shaded bands span the corresponding range between these two extremes (see text). The black dashed
    line is the BLYP $P$--$V$ equation of state of bulk Ice X from
    Ref.~\cite{osmond_ultrahigh_2026_volume_blyp}, shown for comparison.}
    \label{fig:Vwater-explicit}
\end{figure}

For the implicitly confined bilayer (Fig.~\ref{fig:dOO-implicit}), the in-plane and out-of-plane $d_\mathrm{OO}$ separate noticeably across the whole sampled pressure range, with the out-of-plane distance consistently shorter than the in-plane one.
We attribute this to the anisotropic nature of the implicit confinement.
Laterally, the cell relaxes until it matches the applied pressure.
Out of plane, the slit width is held fixed by the wall separation, so the water is compressed to whatever density that width imposes, independently of the lateral pressure.
The out-of-plane O--O distance is therefore set by the walls rather than by the applied pressure, and falls well below its in-plane counterpart.

\FloatBarrier

\section{HSE06 band gaps}

\begin{figure*}
    \centering
    \begin{subfigure}{0.6\textwidth}
        \centering
        \includegraphics[width=\textwidth]{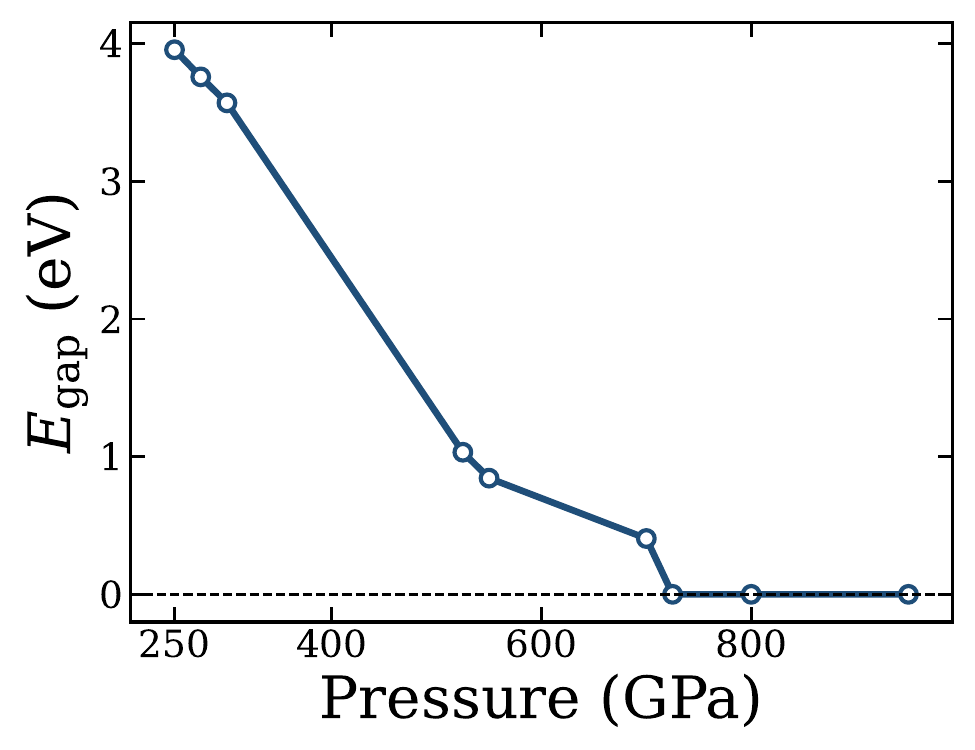}
    \end{subfigure}
    \hfill
    \begin{subfigure}{0.59\textwidth}
        \centering
        \includegraphics[width=\textwidth]{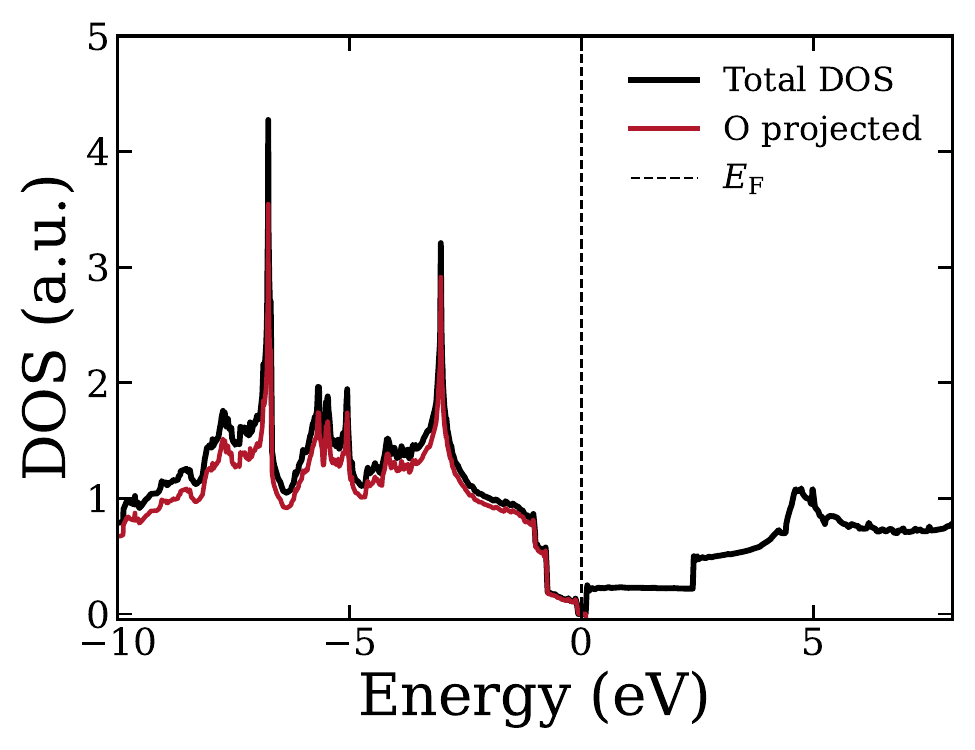}
    \end{subfigure}
    \caption{\textbf{HSE06 electronic structure of the implicitly confined a-Rh-BI phase.} \\
    (Top) Band gap $E_{\mathrm{gap}}$ as a function of pressure. For the 2D ice phase the gap
    closes at approximately $725$~GPa, compared to $500$~GPa at the
    PBE level. (Bottom) Total and O-projected density of states at
    $725$~GPa, where above the Fermi level $E_F$ the H-projected states dominate. Energies are given relative to $E_F$
    (dashed line).}
    \label{fig:hse06}
\end{figure*}

We additionally provide hybrid-functional (HSE06) calculations of the
electronic structure of the implicitly confined a-Rh-BI phase. As expected, the
metallisation pressure shifts upwards relative to PBE, from
$500$~GPa to approximately $725$~GPa
(Fig.~\ref{fig:hse06}a), a shift which is also expected for bulk
high-pressure ice phases. In the hybrid calculations
we employ the intermediate basis set of
FHI-aims. 
As shown in the GGA convergence tests below
(Fig.~\ref{fig:basis-convergence}), at finite gaps a more complete basis
set tends to correct the band gap downwards, with the intermediate basis
becoming increasingly accurate at higher pressures.

Moreover, the hallmark features of the DOS --- H-dominated conduction
states and O-dominated valence states --- remain the same as in the
GGA calculations (Fig.~\ref{fig:hse06}b).

\FloatBarrier

\section{Valence-band character at high pressure}
\begin{figure}[htbp]
    \centering
    \includegraphics[width=0.7\columnwidth]{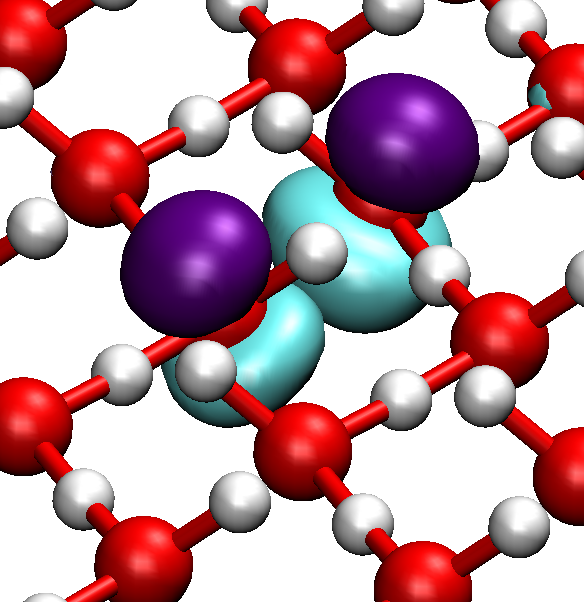}
    \caption{\textbf{Maximally localised valence Wannier functions at high pressure.} \\
    Valence-band MLWFs of the implicitly confined a-Rh-BI phase at 500~GPa, constructed from a fixed energy window reaching down to $-2.5$~eV, spanning the valence states involved in the gap closure. For clarity, only one of the two sheets is shown, and only two Wannier functions are displayed, each located on one oxygen atom. The two isosurface colours indicate different phase signs of the MLWF. The orbitals show a clear 2p structure, consistent with the oxygen-2p lone-pair character of the HOMO in ambient ice and liquid water.}
    \label{fig:wannier-valence}
\end{figure}

\begin{figure*}[htbp]
    \centering
    \includegraphics[width=\textwidth]{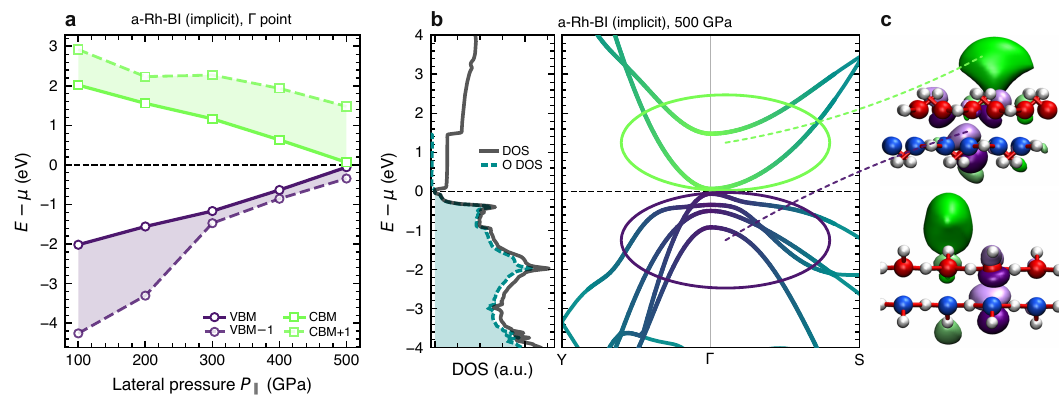}
    \caption{\textbf{Band structure, band-edge splitting and Wannier functions of implicitly confined a-Rh-BI.}
    \\
    \textbf{a)} Valence band maximum (VBM) and its neighbour (VBM$-1$), and conduction
    band minimum (CBM) and its neighbour (CBM$+1$), at the $\Gamma$~point for the
    implicitly confined a-Rh-BI phase as a function of lateral pressure. Shaded areas
    show the VBM/VBM$-1$ and CBM/CBM$+1$ splitting. For the pressure points shown here,
    both the VBM and CBM are always located at the $\Gamma$~point.
    \textbf{b)} Left: total and oxygen-projected electronic DOS for the implicitly
    confined a-Rh-BI phase at gap closure ($500~\mathrm{GPa}$) around the Fermi level
    $E_F$. Right: corresponding band structure, where gap closure occurs at the
    $\Gamma$~point. Green and purple ellipses mark the conduction and valence bands
    around $\Gamma$ whose Wannier functions are shown in c).
    \textbf{c)} Two side views, rotated $90^\circ$ relative to each other, of the
    a-Rh-BI phase at gap closure, showing the maximally-localised Wannier functions
    of the bands within $\pm2.5~\mathrm{eV}$ of $E_F$. White spheres represent
    hydrogen atoms. Blue and red spheres represent oxygen atoms in different sheets.
    Dark and light green lobes correspond to the conduction band states around
    $\Gamma$, while dark and light purple lobes correspond to the valence band states,
    with different shading indicating opposite phase sign.}
    \label{fig:wannier-vbm-cbm-p-evolution}
\end{figure*}

To further characterise the states at the top of the valence band, we compute maximally
localised Wannier functions (MLWFs) for the a-Rh-BI phase discussed in the main
text at 500~GPa, using a fixed energy window reaching down
to $-2.5$~eV. As shown in Fig.~\ref{fig:wannier-valence}, the resulting valence
orbitals are centred on the oxygen atoms and exhibit a clear 2p-like 
structure, with lobes directed out of the plane of the sheet and away from the
covalent O--H bonds. Remarkably, the valence states retain the O~2p
lone-pair character familiar from low-pressure ice \cite{hexdensity} and ambient liquid water. This stands in sharp contrast to bulk ice at high pressure,
where hydrogen-bond symmetrisation often produces a four-fold-coordinated,
symmetric bonding environment in which the lone pairs are absorbed into
bond-aligned, sp$^3$-like orbitals and the low-pressure orbital structure is
lost \cite{icexwannier}. In our system, the reduced dimensionality yields a geometrically constrained hydrogen-bond network, which in turn leaves no bonding partners along the lone-pair directions. The lattice geometry prevents these states from rehybridising into bond-directed combinations, and the lone-pair structure is preserved. Instead, compression progressively localises these states and raises their energy, driving the premature closure of the band gap under implicit confinement. Moreover, in this setup the hydrogen-dominated conduction bands delocalise out of the plane into the vacuum region, as seen in the Wannier functions in Fig.~\ref{fig:wannier-vbm-cbm-p-evolution}. This delocalisation is an artefact of the implicit confining potential commonly used in the confined-ice literature.

However, the lone-pair structure at the top of the valence band is similarly preserved in the explicit diamond anvil setup, as shown by the wave functions at the $\Gamma$~point in Fig.~4 of the main text and the projected density of states in Fig.~\ref{fig:dos-diamond-anvil}. The top of the valence band consists of pure oxygen states, so this feature survives under realistic confinement.

\FloatBarrier

\section{Convergence tests}
\begin{figure}[htbp]
    \centering
    \includegraphics[width=\columnwidth]{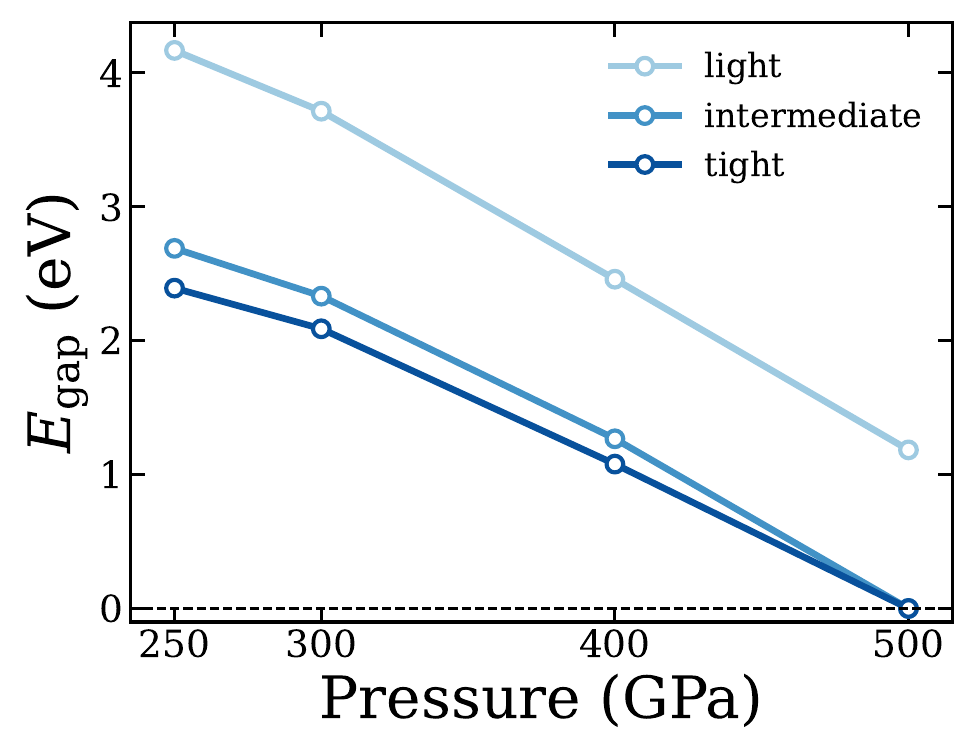}
    \caption{\textbf{Basis set convergence of the band gap.} \\ Band gap
    $E_{\mathrm{gap}}$ of the implicitly confined a-Rh-BI phase as a function of pressure,
    computed with the light, intermediate, and tight species defaults
    of FHI-aims. Here the gaps are reported at the GGA level of theory, employing the PBE functional.
    The intermediate basis set closely tracks the tight
    one, with the agreement improving at higher pressures, whereas the
    light basis set systematically overestimates the gap.}
    \label{fig:basis-convergence}
\end{figure}

We perform convergence test on the implicitly confined a-Rh-BI phase with respect to
the basis set, comparing the light, intermediate, and tight
species defaults of FHI-aims across pressures from $250$ to
$500$~GPa (Fig.~\ref{fig:basis-convergence}). We find that with
increasing pressure the intermediate basis set approaches the tight
one increasingly closely, while the light basis set substantially
overestimates the gap throughout. For computational efficiency, we
therefore employ the intermediate basis set in all calculations. 

Moreover, we test whether the GTH pseudopotentials employed in the CP2K calculations, which we use both for data generation and for the enthalpy ranking of the ice phases, remain fit for purpose at the high pressures considered here. To this end, we select eight representative structures at both $200$ and $400$~GPa, lying within $100$~meV per water molecule of the most stable structure, and recompute their ranking with the all-electron code FHI-aims. The all-electron ranking fully reproduces the CP2K ranking at both pressures, see Table \ref{tab:pp_check}. We therefore conclude that the GTH pseudopotentials are fit for purpose at several hundred GPa.

\begin{table}[ht]
\centering
\caption{Pseudopotential check of the RSS enthalpy ranking. Relative enthalpies $\Delta H$ (meV/H$_2$O)
with respect to the lowest structure at each pressure, computed with revPBE-D3 in CP2K
(GTH pseudopotentials, TZV2P basis) and in all-electron FHI-aims (tight basis) on identical cells
and $k$-grids.}
\label{tab:pp_check}
\begin{tabular}{@{}cccrrr@{}}
\toprule
\multicolumn{2}{c}{Rank} & & \multicolumn{3}{c}{$\Delta H$ (meV/H$_2$O)} \\
\cmidrule(r){1-2}\cmidrule(l){4-6}
CP2K & AIMS & Space group & CP2K & AIMS & AIMS$-$CP2K \\
\midrule
\multicolumn{6}{@{}l}{\textit{200 GPa}} \\
1 & 1 & $C2/m$       &  0.0 &  0.0 &  0.0 \\
2 & 2 & $C2/m$       &  3.5 &  0.4 & $-3.0$ \\
3 & 3 & $P\bar{1}$   &  9.6 &  5.6 & $-4.0$ \\
4 & 4 & $Pc$         & 17.0 & 13.8 & $-3.2$ \\
5 & 5 & $P\bar{1}$   & 29.9 & 33.7 & $+3.8$ \\
6 & 6 & $P2/m$       & 49.0 & 53.6 & $+4.6$ \\
7 & 7 & $P1$         & 58.5 & 60.5 & $+2.0$ \\
8 & 8 & $P2/c$       & 60.1 & 61.2 & $+1.1$ \\
\midrule
\multicolumn{6}{@{}l}{\textit{400 GPa}} \\
1 & 1 & $Pbcm$       &  0.0 &  0.0 &  0.0 \\
2 & 2 & $P\bar{1}$   & 57.9 & 58.9 & $+1.0$ \\
3 & 3 & $P\bar{1}$   & 61.5 & 63.1 & $+1.6$ \\
4 & 4 & $P\bar{1}$   & 67.8 & 69.8 & $+2.1$ \\
5 & 5 & $P\bar{1}$   & 77.6 & 80.0 & $+2.4$ \\
6 & 6 & $P\bar{1}$   & 77.8 & 82.1 & $+4.3$ \\
7 & 7 & $P2_1/c$     & 91.9 & 93.8 & $+1.9$ \\
8 & 8 & $P\bar{1}$   & 95.5 & 95.1 & $-0.4$ \\
\bottomrule
\end{tabular}
\end{table}

\FloatBarrier
\newpage
\FloatBarrier

\bibliography{references}